%% file: icde24-main.tex
\documentclass[conference]{IEEEtran}
\IEEEoverridecommandlockouts
\usepackage{comment}

\usepackage{url}
\usepackage{nth}
\usepackage{amssymb}
\usepackage{algorithm}
\usepackage{algpseudocode}
\usepackage{epsfig}
\usepackage{threeparttable}
\usepackage{footmisc}
\usepackage{hyperref}
\usepackage{paralist}
\usepackage{wrapfig}
\usepackage{enumitem}
\usepackage{subfig}
\usepackage[table]{xcolor}
\usepackage{color}
\usepackage{multirow}
\newcommand{\forceindent}{\leavevmode{\parindent=1em\indent}}
\usepackage{graphicx}

\newcommand{\sharma}[2]{\noindent\textcolor{red}{~#1~}\textcolor{red}{\textbf{(Sharma):}~\textcolor{blue}{#2} \\}} %for displaying Sharma's comments
\def\BibTeX{{\rm B\kern-.05em{\sc i\kern-.025em b}\kern-.08em
    T\kern-.1667em\lower.7ex\hbox{E}\kern-.125emX}}

\begin{document}

\title{\texttt{\underline{MR}-QP:} Scaling Graph Query Processing: Map/Reduce to the Rescue}
\title{\texttt{\underline{MR}-\underline{GQP}:} Leveraging \underline{M}ap/\underline{R}educe to Scale \underline{G}raph \underline{Q}uery \underline{P}rocessing}
\title{Scaling Graph Query Processing Using Partitioning, Focused Exploration, and Load Balancing}
\title{Scaling Graph Query Processing: Focused Exploration on Partitions Using Load Balancing}

\author{\IEEEauthorblockN{Harshit Modi}
\IEEEauthorblockA{\textit{IT Lab \& CSE Department} \\
\textit{University of Texas at Arlington}\\
Arlington, Texas, USA \\
harshitashokkum.modi@mavs.uta.edu}
\and
\IEEEauthorblockN{Abhishek Santra}
\IEEEauthorblockA{\textit{IT Lab \& CSE Department} \\
\textit{University of Texas at Arlington}\\
Arlington, Texas, USA \\
abhishek.santra@mavs.uta.edu}
\and
\IEEEauthorblockN{Sharma Chakravarthy}
\IEEEauthorblockA{\textit{IT Lab \& CSE Department} \\
\textit{University of Texas at Arlington}\\
Arlington, Texas, USA \\
sharmac@cse.uta.edu}
}

\maketitle

\begin{abstract}
    \input{texFiles/abstract}

\end{abstract}

\input{texFiles/introduction.tex}
\input{texFiles/related-work.tex}

\input{texFiles/graph-query-processing.tex}
\input{texFiles/mr-algorithms.tex}

\input{texFiles/addressing-challenges-correctness.tex}

\input{texFiles/experiments.tex}
\input{texFiles/conclusions.tex}
\input{texFiles/acknowledgement.tex}

\clearpage

\bibliographystyle{IEEEtran}
\bibliography{./bibliography/harshit,./bibliography/somu_research,./bibliography/itlabPublications,./bibliography/itlabTheses,./bibliography/santraResearch,./bibliography/itlabPublications.original}

\end{document}

%% file: texFiles/abstract.tex
The focus of this paper is on query processing and not search (approximate or exact). The goal is to find \textit{all} or k answers and the expressiveness of a query (as opposed to a search expression which is typically simpler) is considered important.
Although SQL query processing has matured over the last several decades, query processing over graph data has not received the same attention. 
%This could be partly because there were no industrial-strength DBMSs that used graph representation. 
Only recently, systems, such as Neo4J, %%ArangoDB, OrientDB, Amazon Neptune,  
and others have started using graphs for data representation on a large scale. Their approaches and algorithms remain proprietary.
%%Notwithstanding a lot of effort on \textit{searching} graphs for small subgraphs, querying has received little attention. Knowledge graphs and other large social networks warrant querying them for all instances of a desired graph. In addition, querying provides an expressive alternative to look for intended subgraph. \\
%This work is different from graph mining for exact and similar substructures. Due to the ubiquitous use of graphs and their growing sizes, query processing on them has become not only relevant, but also important.

\forceindent In this paper, we address the problem of answering graph queries and their efficient processing on arbitrary-sized graphs. We propose the partitioning of graphs and the use of the distributed paradigm to achieve efficiency and scalability. Our approach takes a subgraph as query input including wild cards and comparison operations. Our partitioned approach leverages focused exploration and distributed processing to achieve the desired parallelism and improve the response time. We also present heuristics to reduce the number of partitions loaded for query processing. We present our partitioning algorithm, distributed processing architecture used, and load balancing for computing query answers. We establish the correctness of the approach in the presence of partitions, compare heuristics for load balancing, and extensive experimental results over large synthetic and real-world graph data sets. 
%%We highlight the research challenges encountered due to \textit{independent} processing of \textit{arbitrary partitions} and producing all (or top k) correct results. Efficiency of our approach with respect to the numbers of mappers and reducers are also analyzed.
%\vspace{-10pt}

%% file: texFiles/introduction.tex
\section{Motivation}
\label{sec:motivation}

As the amount of structured data and application characteristics are increasing, graphs used for modeling them are becoming larger and they cannot be stored in main memory or use a single processor for bounding the response time. For example, Facebook has around 2.3 billion active users \cite{fbusers}, computer science bibliography website, DBLP, has over 5 million records of conferences/workshop papers and journal articles \cite{dblpstats}, and Internet Movie Database (IMDb) contains over 360 million data items with information of 123 million people and 6.5 million titles \cite{imdbstats}. This brings significant challenges in querying large graphs.

\forceindent We propose a partitioned approach along with leveraging a distributed processing framework for processing queries on large graphs. When query evaluation is done on partitioned graphs, answers may span multiple partitions which brings new challenges to make sure: \textit{i)} all answers (or k answers) are computed (completeness) and \textit{ii)} all computed answers are correct (soundness). Our approach allows us to match partition sizes to the processor capability as well as control the amount of parallel processing. For this approach, we also \textbf{propose a number of heuristics for choosing which and how many partitions to evaluate after each iteration to minimize the total number of partitions loaded and to reduce total response time} (Load balancing). We also formulate a load factor to assess the efficacy of a heuristic used. Based on the availability/affordability of resources, our approach has the \textbf{ability to adjust to specified/affordable amount of parallel processing}. 
%So we can use as many partitions and processors as needed to process in parallel. 
In the ideal case, we can match the number of partitions with the corresponding number of resources in each iteration for maximum parallelism and the best response time.
%But in case of limited resources, our approach still works by processing more than one partition on a single machine.

%%SC: \abhi{9/17}{SC: clarified a bit: We may have to re-phrase/formalize the problem statement. I have taken a crack. Please check. Also, does MR-GQP as input require a query "plan" to be given by the user, or does the system generate it, may be using a call to QP-Subdue? We may have to highlight that.}

\textbf{Problem Statement}: 
The problem is to develop a distributed processing framework to process graph queries on very large graphs with reasonable or controlled response time. Specifically, given a large graph $G$ and available resources (in terms of total number of processors/cores), develop a graph partition strategy $P$, a set of partition loading heuristics $H$, and a distributed algorithm that can efficiently and correctly (satisfy soundness and completeness) process any \textit{expressive} graph query $Q$. %(with relational comparison operators ($>$, $>=$, $<$, $<=$, $=$, $<>$), range operators (Between, IN), wild cards (multiple (\%) and single (\_) character match), and Boolean operators AND, OR.)

\forceindent The sub-problems of this are: \textbf{i)} expressive query specification, \textbf{ii)} developing partitioning strategies, \textbf{iii)} distributed processing algorithms to process partitions using proposed heuristics and obtaining all (or k) answers, and \textbf{iv)} establishing soundness and completeness of results. 

\subsection{\textbf{Why Distributed Processing?}}

Having experimented with parallel processing approaches~\cite{msThesis/Bodra16}, we believe an established distributed processing paradigm is better suited for this work. Then there is the question of which one to use and why among many choices, such as Spark, Hadoop, Pregel, etc. We want to assert that our work is not for \textit{measuring absolute response time}. Instead, our goal is to develop a scalable query processing approach using \textbf{a} paradigm and validate it analytically and experimentally. Any choice of framework is neutral to our focus and the use of Hadoop and Map/Reduce is based on availability and convenience. It is true that some algorithms can benefit from, say, Spark. But our algorithms do not have such in-memory storage requirements across iterations to take advantage of, say Spark. Finally, our approach is architecture agnostic and the algorithms do not change much. Only the absolute values will change and trends are likely to remain the same.

\forceindent We have chosen the publicly available Hadoop Map/Reduce framework as a distributed paradigm that has been used to scale computations horizontally to accommodate very large data sizes. This paradigm has been successfully used for a number of graph problems, such as search~\cite{indexingArticle} and approximate matches~\cite{giugno2002graphgrep,tong2007fast}, counting triangles~\cite{CIKM/ArifuzzamanKM13}, etc.  Apache Hadoop, which is an open-source implementation of Map/Reduce, has been used extensively. The choice of Map/Reduce instead of traditional parallel processing is obvious. Partitioning still remains the same. However, using multiple processors, scheduling them, passing data among processors, and loading needs to be done by the query processing system. In addition, aggregating partial results and bookkeeping for that is quite complicated. On top of this, failure recovery of processors needs to be handled by the query processing system. This was experienced in our previous work ~\cite{DaWak/das2016plan,bodra2018query} which led to the choice of the Map/Reduce paradigm.

\forceindent Hadoop framework offers a number of advantages and also some challenges. The main challenge is to cast the algorithm into Map/Reduce framework and figure out how to exploit the shuffle and merging. Correctness of the algorithm, in the presence of partitioning, is also to be addressed. However, the Hadoop framework handles heterogeneous processors, different size partitions to match processing resources, choice of number of processors for parallelism, and robust execution (fault tolerance) in the presence of failures.

\begin{comment}
A Map/Reduce framework is composed of the following steps:

\begin{itemize}
    \item \textbf{map phase}: Processes a key/value pair to generate a set of intermediate key/value pairs
    \item \textbf{optional combiner}: to aggregate mapper output to reduce shuffling and generates a set of key/value list elements
    \item \textbf{shuffle phase}: Distribute the data based on key partitions such that all keys in a partition are sent to the same reducer
    \item \textbf{reduce phase}: Merges all intermediate values associated with the same intermediate key and invokes a user defined reduce function.
\end{itemize}
\end{comment}

If we implement our algorithm using traditional multi-processor implementation, extensive coding is needed for inter-process communication and bookkeeping. In our implementation, the map, shuffle, and reduce paradigm are used to reduce bookkeeping and inter-process communication.  Moreover, Map/Reduce has also been used for partitioning the graph based on the number of resources available.

\subsection{Choice of Data Sets}

We have chosen IMDb and DBLP data sets for experimental purposes for a reason. Although the data sets ate not extremely large, they allow us to  verify query results comparing them with their database versions and writing corresponding SQL queries. However, the approach is general and scales to any number of Mappers/Reducers depending on the data size. This also helps us demonstrate the expressiveness of our query input with SQL in contrast to other work.

\vspace{5pt}

The contributions of this paper are:
\begin{itemize}
\item \textbf{Scalable approach to graph querying} on arbitrary sized graphs -- directed, undirected, labeled.
\item \textbf{Support expressive graph queries} - handle relational comparison operators $(>, >=, <, <=, =, <>)$, range operators (Between, IN), wild cards (multiple character match (\%), single character match (\_), and  Boolean operators AND and OR.
\item A \textbf{variable-size range partitioning Map/Reduce} algorithm that takes a single pass to create graph partitions of different sizes and their adjacency lists.
\item \textbf{Several heuristics} to optimize the loading of partitions for efficient query evaluation and their analysis.
\item \textbf{Map/Reduce} algorithm for leveraging distributed processing for querying.
\item \textbf{Extensive experimental analysis} on large real-world and synthetic data sets using a broad range of queries to validate the proposed heuristics and the approach.
\end{itemize}

%% file: texFiles/related-work.tex
\section{Related Work}
\label{sec:related-work}

Although there has been a lot of effort on graph search \cite{giugno2002graphgrep,yan2004graph,cheng2007fg,cheng2009efficient,tong2007fast} and this paper is about graph querying, we limit ourselves to discussing earlier querying languages~\cite{angles2017foundations} and approaches for graphs.

%%\subsection{Query languages for graphs}
%%Graph querying has been studied for specific graph data models by formulating different query languages\cite{angles2017foundations}.
%%\subsection{SPARQL}

\forceindent SPARQL query language\cite{prud2008sparql} is a W3C recommendation for querying RDF (Resource Description Framework) graph data. RDF is a directed, labeled graph data format for representing information on the Web. It describes a graph by a set of triples, each of which describes a pair (attribute and value) or an interconnection between two nodes. The SPARQL query language works primarily through a primitive triple pattern matching technique with simple constraints on query nodes and edges.

\begin{comment}
\begin{figure}
\vspace{-20pt}
    \centering
    \includegraphics[width=1\columnwidth]{figures/sparql.png}
    \caption{SPARQL query pattern}
    \label{fig:sparql}
\end{figure}
\end{comment}

\forceindent Cypher\cite{cypher} is Neo4j's \cite{neo4j} graph query language, that allows for expressive and efficient data querying in a property graph. Patterns are expressed syntactically following a ``pictorial" intuition to encode nodes and edges with arrows between them. 

\begin{comment}

The query ``Find friend(s) of bob" would be written in Cypher as:

%\vspace{-5pt}
\footnotesize
{
\begin{verbatim}
MATCH (you {name:"Bob"})-[:FRIEND]->(yourFriends) 
RETURN you, yourFriends
\end{verbatim}
%\vspace{-5pt}
}

\normalsize

\forceindent MATCH clause specifies the pattern. Nodes are written inside ``( )" brackets and edges inside ``[ ]" brackets. The RETURN clause can be used to project the output variables. 

\end{comment}

However, above mentioned applications require the query to be represented in a \textit{specific} language model. Hence writing structured queries requires extensive experience in query language and the data model, as well as a good understanding of the particular data set. \textit{Our approach takes a query as a graph and finds all instances of it in a very large graph.}

It is to be noted that there is some work addressing subgraph enumeration (\cite{DBLP:journals/corr/abs-2103-14294,DBLP:journals/pvldb/AmmarMSJ18}) where triangles, cliques, diamonds, etc. are identified, which is not the same problem as query evaluation, which is the focus of the paper. Moreover, such work will not be able to process the large class of expressive queries that have been taken care of in the proposed approach. 

\forceindent In a nutshell, as compared to earlier work on graph query processing, this work differs in a number of ways. Scalability is being addressed in a general-purpose manner using a different paradigm. This work extends upon our earlier work on partitioned non-Map/Reduce approach for query processing~\cite{msThesis/Bodra16,qpsubdue,DaWak/das2016plan}. This paper handles more expressive query representations. Also, the heuristics proposed in this paper are different and relevant to the framework used. Finally, there are no restrictions on the number of entities or labels that can be specified as part of a query. We can find all (or k) exact matches in very large graphs. Using the approach proposed in this paper, ``query relaxation"~\cite{IEEE/Gaasterland97} can be easily incorporated as the expression computed is similar to an SQL query, and edge labels are similar to attribute values. In this approach, answers can also be ranked to output results based on a given threshold.

%% file: texFiles/graph-query-processing.tex
\section{Large Graph Query Processing Challenges}
\label{sec:graph-query-processing}

Three major challenges for querying large graphs are: \textit{i)} how to partition the graph to minimize computation overhead for answers spanning multiple partitions, \textit{ii)} how to preserve the correctness of the results in the presence of partitions, and \textit{iii)} what heuristics are appropriate for improving computation efficiency. These issues arise due to graph partitioning, the need to load and process partitions independently (and may be more than once), and the bookkeeping required to compute correct and complete results.

%%\subsection{Need for Variable-size Partitioning}

\vspace{5pt}

\noindent \textbf{Need for Variable-size Partitioning:} There are several ways to partition a graph and this has been researched extensively (e.g., KaHIP~\cite{KaHiP2011,KaHipThesis2013} and Metis~\cite{Metis1998}.)
%In this paper, we assume that the entire information needed (graph, adjacency list of the graph) cannot be stored in the memory of a single machine. Hence partitioning the graph among multiple processors is needed where each partition can be held in the memory of a single processor. 
Existing approaches, typically, make multiple passes over the input or produce equal-sized partitions, thus affecting the overall performance. Moreover, they are unable to match partition sizes based on the processing resource capabilities. Variability of processing resources is common in Map/Reduce framework. 

\begin{figure}[h]
\centering
\vspace{-15pt}
\includegraphics[width=0.5\columnwidth]{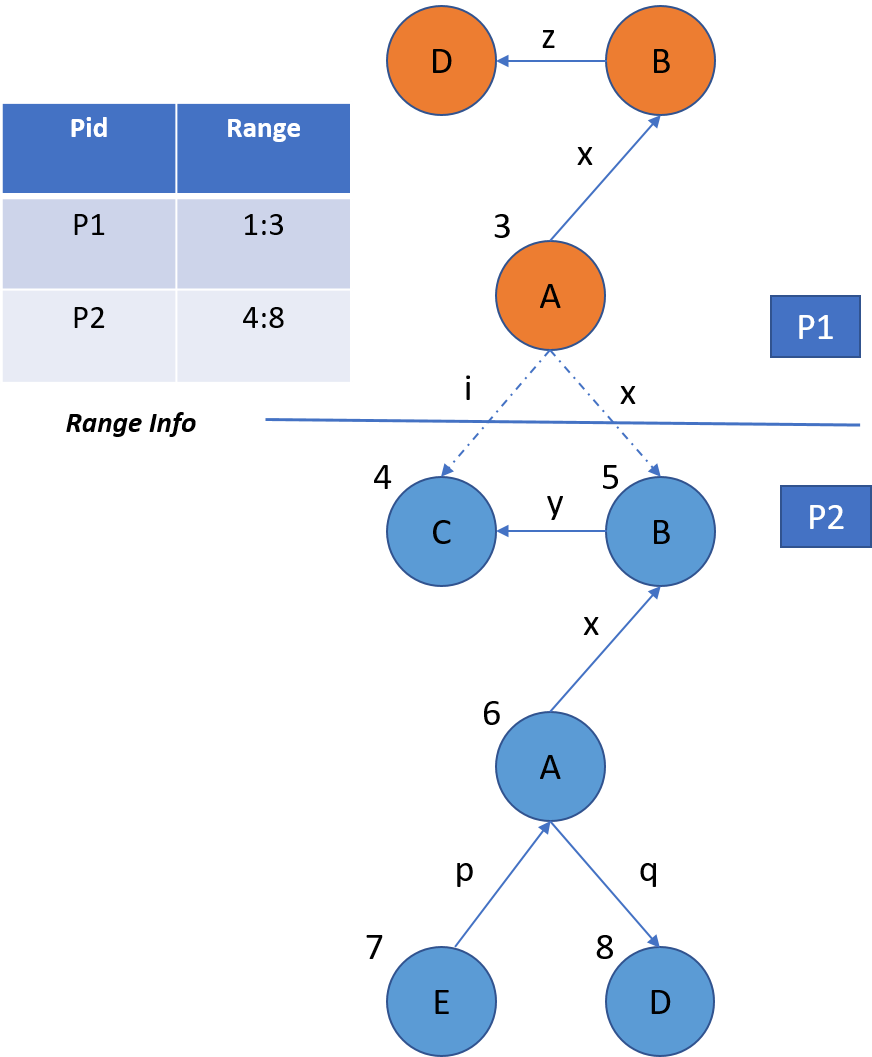}
%%includegraphics[width=0.90\columnwidth]{figures/Picture1.png}
\caption{Range Partitioning}
\vspace{-15pt}
\label{fig:range_partitioning}
\end{figure}

\begin{table}[h]
\caption*{\textbf{\small Adjacency Lists}}
\vspace{-5pt}
\begin{minipage}{.35\columnwidth}
\footnotesize
  \centering
    \begin{tabular}{|p{0.3cm}|p{1.5cm}|}
        \hline
            \textbf{Vid} & \textbf{Adj List } \\
        \hline
        \hline
        1 & (z,2,B,1,D)\\
        \hline
        2 & (z,2,B,1,D), (x,3,A,2,B) \\
        \hline
        3 & (x,3,A,2,B), (i,3,A,4,C), (x,3,A.5,B) \\
        \hline
    \end{tabular}
    \caption{\textbf{\small Partition 1}}
    \label{fig:Partition 1 Adjacency List}    
\end{minipage}%
\begin{minipage}{.45\columnwidth}
  \footnotesize
  \centering
    \begin{tabular}{|p{0.3cm}|p{4.3cm}|}
        \hline
            \textbf{Vid} & \textbf{Adj List } \\
        \hline
        \hline
        4 & (i,3,A,4,C), (y,5,B,4,C) \\
        \hline
        5 & (x,3,A,5,B), (y,5,B,4,C), (x,6,A,5,B) \\
        \hline
        6 & (x,6,A,5,B), (p,7,E,6,A), (q,6,A,8,D) \\
        \hline
        7 & (p,7,E,6,A) \\
        \hline
        8 & (q,6,A,8,D) \\
        \hline
    \end{tabular}
    \caption{\textbf{\small Partition 2}}
    \label{fig:Partition 2 Adjacency List}
\end{minipage}
\vspace{-15pt}
\end{table}

\begin{comment}

\begin{table}[h]
\centering
\footnotesize
%\vspace{-30pt}
    \begin{tabular}{|p{0.3cm}|p{3cm}|}
        \hline
            \textbf{Vid} & \textbf{Adj List } \\
        \hline
        \hline
        4 & (i,3,A,4,C), (y,5,B,4,C) \\
        \hline
        5 & (x,3,A.5,B), (y,5,B,4,C), (x,6,A,5,B) \\
        \hline
        6 & (x,6,A,5,B), (p,7,E,6,A), (q,6,A,8,D) \\
        \hline
        7 & (p,7,E,6,A) \\
        \hline
        8 & (q,6,A,8,D) \\
        \hline
    \end{tabular}
    \caption{\textbf{Partition 2 Adjacency List}}
    \label{fig:Partition 2 Adjacency List}
    \vspace{-20pt}
\end{table}

\end{comment}

\forceindent Instead of arbitrary/random partitioning or partitioning using a power of 2 (as is done in Metis), we use \textit{range partitioning} for its flexibility, superior performance during pre-processing, and effectiveness in substructure discovery \cite{phdThesis/Das17}. We use the contiguous range of vertex ids (a range need not have all numbers in the range) to partition the input graph and create an adjacency list for each partition. In the example given in Figure \ref{fig:range_partitioning}, to partition the graph into two using given range information, we assign the first 3 nodes (vertex ids 1 to 3) to partition 1 and the other 5 nodes (vertex ids 4 to 8) to partition 2. With this, adjacency lists for those two partitions are created as shown in Tables \ref{fig:Partition 1 Adjacency List} and \ref{fig:Partition 2 Adjacency List}.

%%\sharma{2/23/2022}{lots of white space to be recovered; recovered some}

% \sharma{}{same question regarding direction in the adj list}
% \harshit{}{addressed the same}

%\vspace{-20pt}

\forceindent Range partitioning takes \textbf{one pass on the input graph} and is done using Map/Reduce only once and the cost is amortized over processing as many queries on the graph as needed. We also generate meta information during partitioning for computing heuristics to be used during query evaluation. 
%The information collected include query plan start nodes\footnote{Query sizes are significantly smaller compared to the graph/partition and this information can be collected for multiple queries as well.} in each partition and intermediate query plan nodes in each partition, and connected components in each partition. 
In each partition, the information collected include query plan start nodes\footnote{Query sizes are significantly smaller compared to the graph/partition and this information can be collected for multiple queries as well.}, intermediate query plan nodes, and connected components. 
Partitions generated after range partitioning contain partitioned graphs (could be a forest)
%in that partition
and separately its adjacency list. As query expansion on these partitioned graphs needs corresponding adjacency list partition, we need to direct this partition to the processor holding the appropriate adjacency list. Without range information among vertex ids across the adjacency partitions, this operation requires the complete information of \textit{all vertices and their partition adjacency lists}. 
%This information of all vertices will be large and can not be stored in the memory of a single processor. 
Range information allows us to \textbf{direct a partial answer to the correct partition in which the adjacency list of node is stored}. Also, as we use the mapping between partition id and vertex id range stored as \textit{range information}, this gives us \textbf{finer control over the size of the partitions for heterogeneous processing resources}.

%%\subsection{Correctness}

%\abhi{9/17}{We may have to add how does a query look - the representation in a way that the user inputs it. What do you think?}

%\vspace{3pt}

\begin{figure}[H]
    \centering
   \vspace{-10pt}
    \includegraphics[width=\columnwidth]{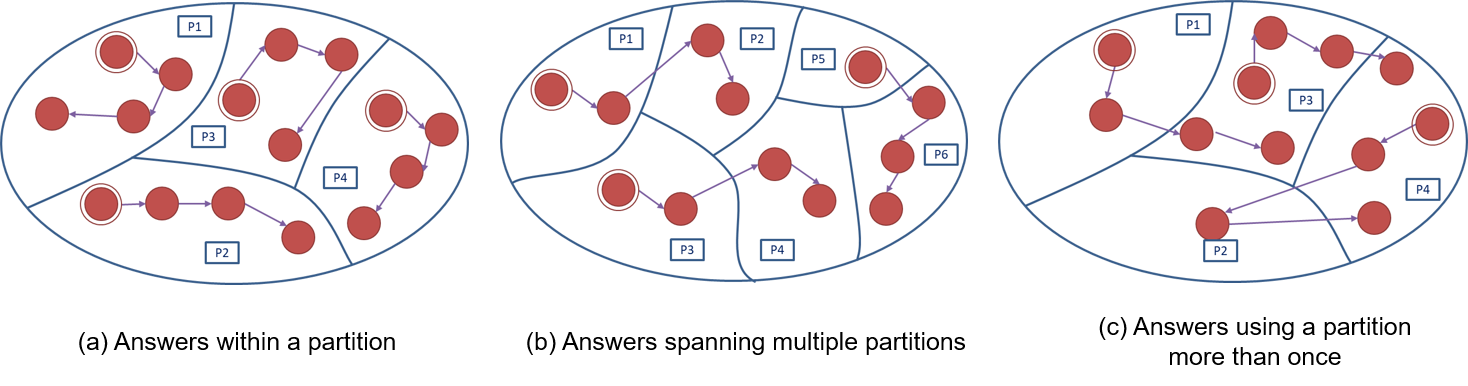}
    \vspace{-20pt}
    \caption{Challenges for Querying Partitioned Graphs}
    \label{fig:3_scenarios}
    \vspace{-10pt}

\end{figure}

\noindent \textbf{Correctness:} As answers to a query can be part of multiple partitions and even come back to an already processed partition, several cases (shown in Figure \ref{fig:3_scenarios}) need to be systematically dealt with for answer correctness.

%%Processing a query on partitioned graph is very different as compared to processing a query on single graph. The three cases that need to be addressed for answering queries correctly are shown in Figure \ref{fig:3_scenarios}.
    
%    \vspace{-10pt}

\begin{enumerate}
    \item All answers are within a single partition (Figure \ref{fig:3_scenarios} (a))
    \item Answers span more than one partition (Figure \ref{fig:3_scenarios} (b))
    \item Answers span one or more  partitions that are already processed (Figure \ref{fig:3_scenarios} (c))
\end{enumerate}

%\abhi{9/17}{Is there a way to highlight in the algorithm that the 3 cases highlighted above have been handled and thus show the correctness of the MR algorithm (algo 2) in general?}

\begin{figure}[H]
\centering
\vspace{-10pt}
\includegraphics[width=0.75\columnwidth]{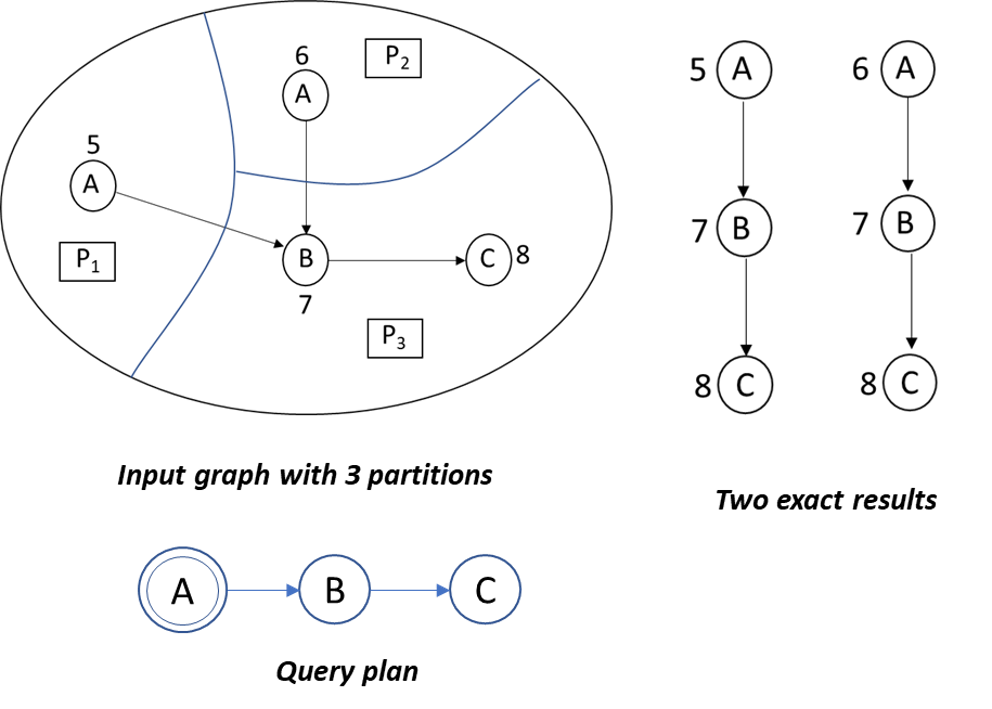}
\vspace{-10pt}
\caption{Overlapping answers}
\vspace{-5pt}
\label{fig:overlap}
\end{figure}

In addition to the above, there are additional challenges to make sure answers are correct in the presence of non-unique \textit{node and edge labels} (node ids are unique.) They are: 
%%Below, we discuss different scenarios where we have to take measures for correctness of query evaluation.

\vspace{3pt}
\noindent\textbf{\underline{Overlapping Answers}:}
Figure \ref{fig:overlap} shows an example where the answer instances overlap. That is, the vertex or edge is part of more than one answer (due to their label.) Here, our query processing should identify all the instances as separate answers for correct results. Note that node B in partition P3 needs to be expanded twice once for each answer to get correct answers.

\begin{comment}

\begin{figure*}[h]
\vspace{-30pt}
\raggedleft
\begin{minipage}{0.50\textwidth}
%\vspace{-50pt}
\includegraphics[width=0.95\columnwidth]{figures/overlap.png}
\caption{Overlapping answers}
\label{fig:overlap}
\vspace{-10pt}
\label{fig:Graph-example}
\end{minipage}%
\raggedright
\begin{minipage}{0.50\textwidth}
\includegraphics[width=\columnwidth]{figures/multi_path_query_plan.png}
\vspace{-20pt}
\caption{Issue with multiple path query plan}
\label{fig:mp_qp}
\vspace{-20pt}
\label{fig:Grapwraph-example}
\end{minipage}%
\end{figure*}

\end{comment}

\begin{figure}[h]
\centering
\vspace{-12pt}
\includegraphics[width=0.9\columnwidth]{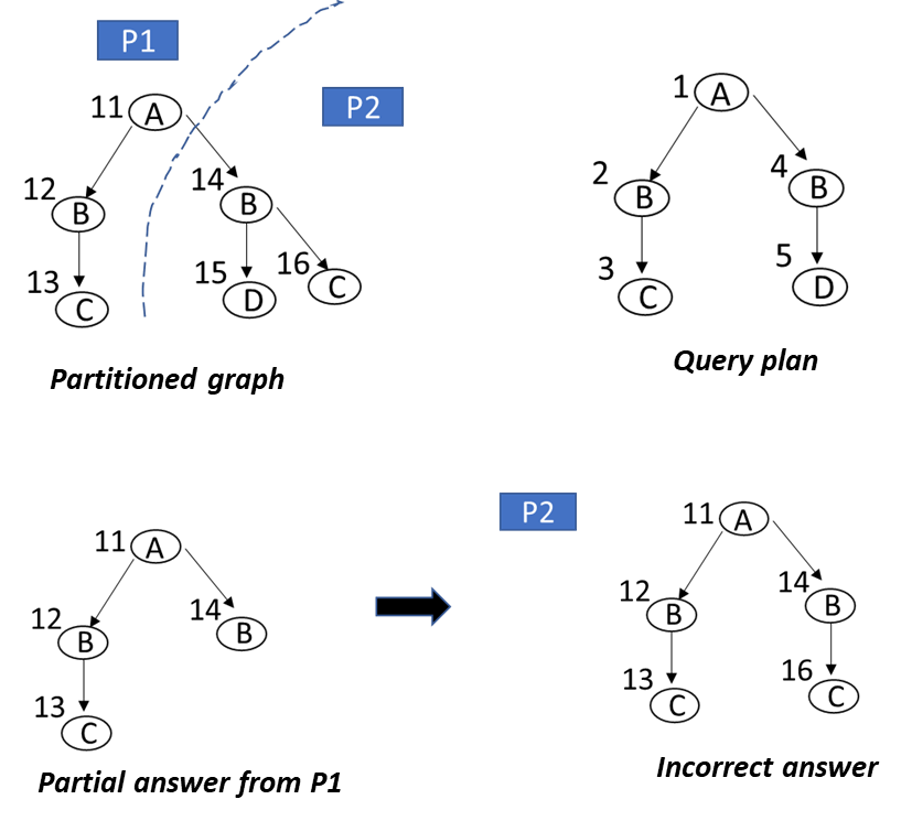}
\vspace{-10pt}
\caption{Issue with multiple path query plan}
\label{fig:mp_qp}
\vspace{-10pt}
\end{figure}

\noindent\textbf{\underline{Multi-Path Query Plans}:}
Based on query characteristics, a query plan can have more than one path from top to bottom. If the plan contains multiple traversal paths, then the complete answers should match the query edges present in all paths. Figure \ref{fig:mp_qp} shows a multiple path query plan and the issues with processing it. The partial answer found in partition P1 continues for node (14, B) in partition P2 and becomes start node in that partition. Then, we load and process all start nodes in partition P2. However, we need to keep track of which query plan node needs to be expanded (as there is another B node in the query plan). If we expand the node (14, B) in P2, thinking it is node (2, B) of query plan, the result will be incorrect (using the edge $B \rightarrow C$ from the query plan). Thus, without proper bookkeeping, we may find incorrect results. \textit{This situation warrants not only keeping track of which node to expand, but how it needs to be expanded with respect to query plan.}

%%sc: need to fix the above para

\begin{figure}
%\vspace{-20pt}
    \centering
    \includegraphics[width=\columnwidth]{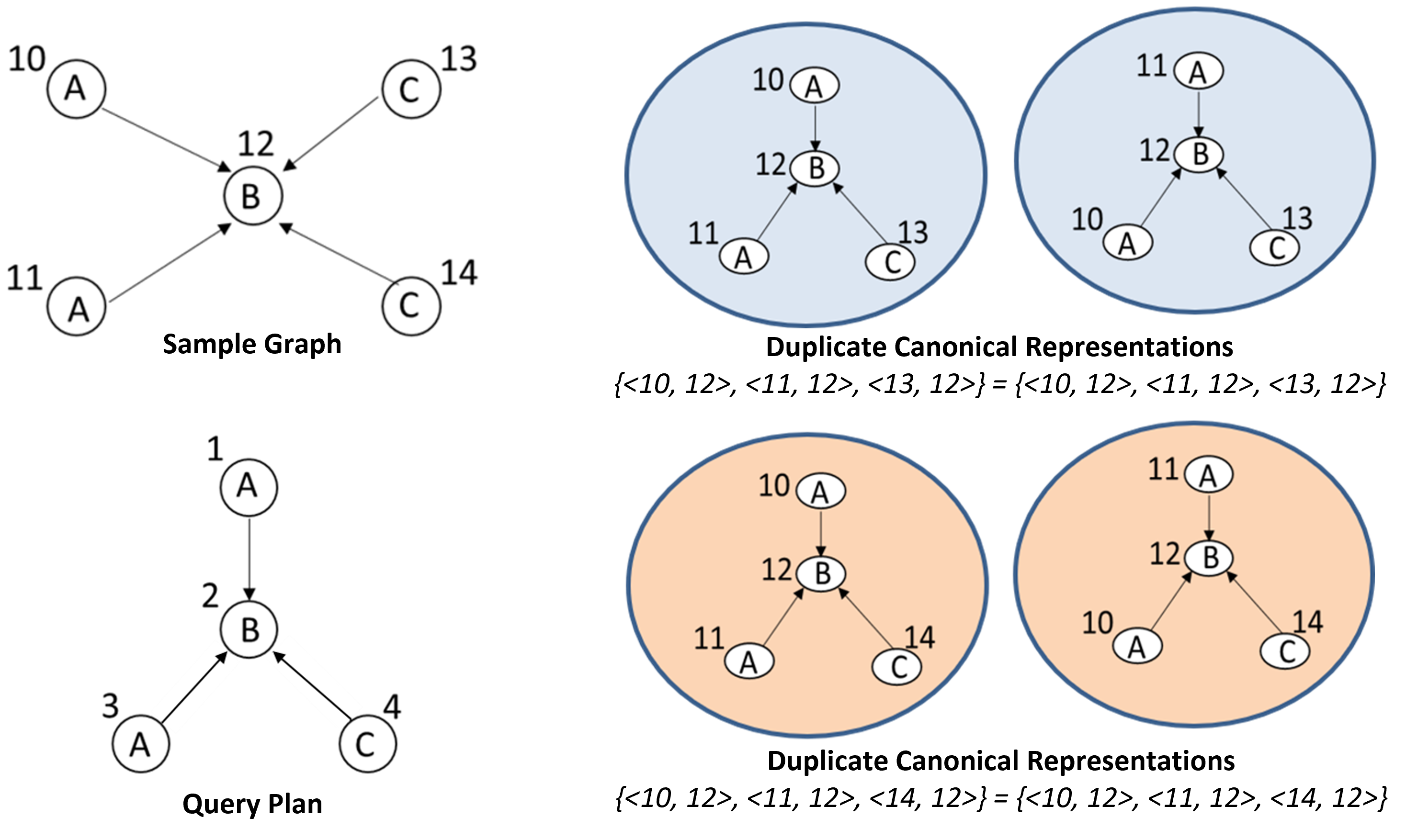}
 %   \vspace{-15pt}
    \caption{Duplicates in Query Expansion}
    \label{fig:duplicates}
    \vspace{-20pt}
\end{figure}

\noindent \textbf{\underline{Duplicates in Query Expansion}:}
%\abhi{3/14/2022}{check}
Independent and one-edge-at-a-time expansion of the graph from query start node and continuing may lead to the generation of \textit{duplicates}. Hence, duplicate elimination~\cite{tkde/DasC18} is needed to find correct answers and reduce the cost of query expansion. Figure~\ref{fig:duplicates} shows an example of how duplicates are formed during query expansion. The circled instances highlighted in same color have the same set of vertex ids and edges between them. In the presence of duplicates, query processing not only uses resources, but also generates incorrect answers. \textit{We use canonical representation to identify duplicates.}

\section{Query Processing Approach and Its Correctness}

Our approach uses the Map/Reduce framework to support querying graphs of arbitrary sizes. We use a parameter k to indicate the number of available resources (in our case Mappers). For a given k, we choose the best k partitions to be processed in parallel in each iteration using heuristics and evaluate a given query plan.

\noindent\textbf{Constrained Query Plan Expansion:}
In our approach, a query plan\footnote{Given a graph query, several plans are possible. Generating a low-cost query plan is discussed in~\cite{DaWak/das2016plan}.} is used to determine the order in which nodes should be expanded to obtain the answer instances that match the given query. Constrained query expansion in a partition starts from initial instances of the start nodes and then we expand all instances one edge at a time, independently according to query plan. As we expand, if the query edge goes out of current partition, we update the corresponding partition with the partial answer instance. We do so until there are no more edges to visit in the query plan (or plans) or no more answers to process in that partition.

\noindent\textbf{Removing Duplicates:}
Duplicates have the same vertex ids and the same connectivity including node labels and edge direction and labels, if present. To remove duplicates we convert the expanded instance to its canonical form. We employ a lexicographic ordering on edge label for canonical form. If there are multiple edges with the same edge label in a substructure, they are ordered on the source vertex label. If the source vertex label is also the same, they are further ordered on the destination vertex label. If edge label, and vertex labels are also identical, then source and destination vertex ids are used for ordering. Canonical forms of duplicates are shown in Figure~\ref{fig:duplicates}.
%\sharma{2/26/2022}{change edge directions of 3,2 and 4,2 in query plan}

%\sharma{3/4/3022}{Add to this figure \{$<10,12>$ $<11,12>$ $<14,12>$\} and \{$<10,12>$ $<11,12>$ $<13,11>$\}}

\begin{comment}
\begin{figure}
    \centering
    \includegraphics[width=0.90\columnwidth]{figures/cano_ordering.png}
    \caption{Canonical ordering of instances}
    \label{fig:cano_ordering}
\end{figure}
\sharma{3/3/2022}{show as canonical representation instead if a graph}

Figure \ref{fig:cano_ordering} shows an example of how lexicographical ordering during expansion handles duplicate elimination. The start node instances (10, A) and (11, A) are expanded according to query plan and as we get duplicate instances (10, 12, 11) and (11, 12, 10), we merge them by canonical ordering.
\end{comment}
% \sharma{}{in the above, what is used for lexi order? node label? as there are no edge labels. the edge directions seem to be not correct. check}

% \sharma{6/2/2020}{can you answer this?}

\noindent \textbf{Routing Partial Answers:}
When all query results are completely inside a single partition, complete answers are computed and stored while processing that partition. When query results span multiple partitions, we compute partial answers in one partition and carry that instance to the appropriate partition which has further query edges. We keep constructing the answers by routing it to appropriate partition and loading additional partitions until the query plan is completely matched. The complete answers are collected when the answer instances have all the edges according to the query plan. When an answer instance comes back to a partition that has already been processed, it is a special case of the above. As we carry instances and process additional partitions, we ensure correctness by loading those partitions as many times as needed.

\noindent \textbf{Handling Overlapping Answers:}
Our system makes sure that our final result set contains all overlapping answers as \textit{separate instances}. We compute query on each initial substructure systematically and if any answer goes out of partition, we carry them as \textit{separate instances} and systematically keep finding complete answers. For the example discussed in Figure \ref{fig:overlap}, the initial instances (5, A) and (6, A) will be carried separately to partition P3 and when we process partition P3, node (7, B) will be used for expanding both instances. Hence, all the overlapping answers come out correctly.

\begin{figure}[H]
\vspace{-10pt}
    \centering
    \includegraphics[width=1.0\columnwidth]{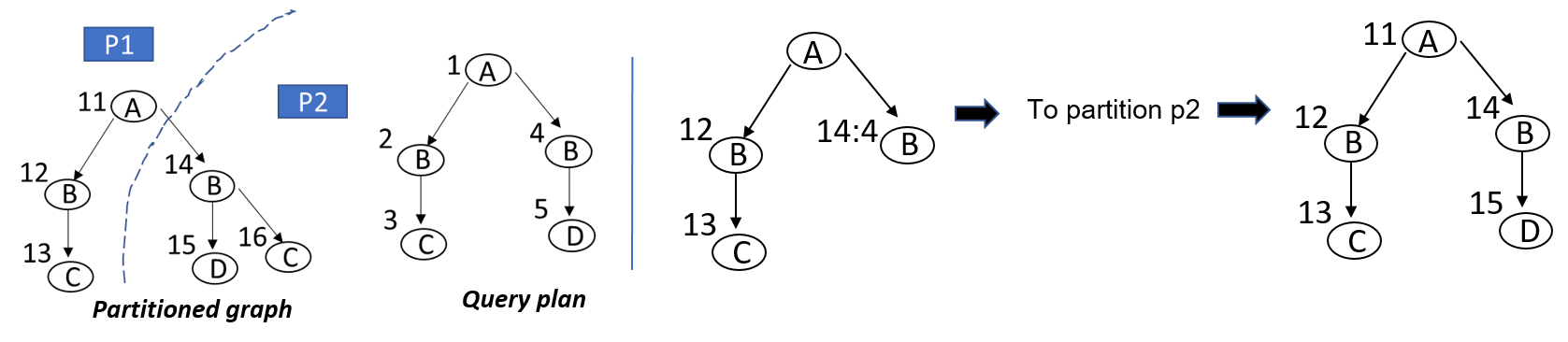}
    %\vspace{-15pt}
    \caption{Marking node for correctness}

    \label{fig:mark_node}
    \vspace{-10pt}
\end{figure}

\begin{figure*}[h]
%\vspace{-25pt}
    \centering
    \includegraphics[width=0.7\textwidth]{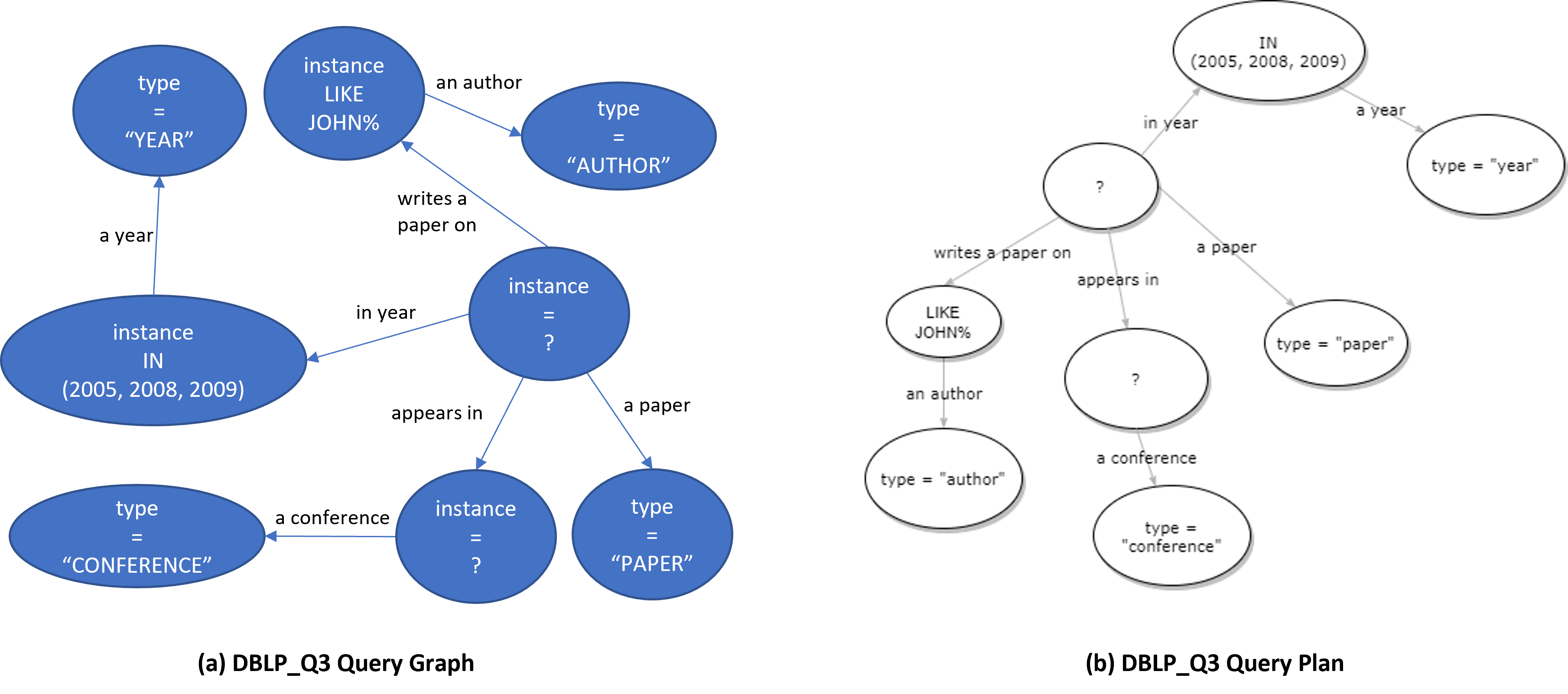}
%    \vspace{-5pt}
    \caption{Query Graph and Plan for DBLP\_Q3: \textit{Find all conference papers along with the conference names that are published in 2005 or 2008 or 2009 by the author whose first name start with `John'}}
    \label{fig:dblp-q3}
 %   \vspace{-20pt}
\end{figure*}

\begin{figure*}[h]
%\vspace{-25pt}
    \centering
    \includegraphics[width=0.7\textwidth]{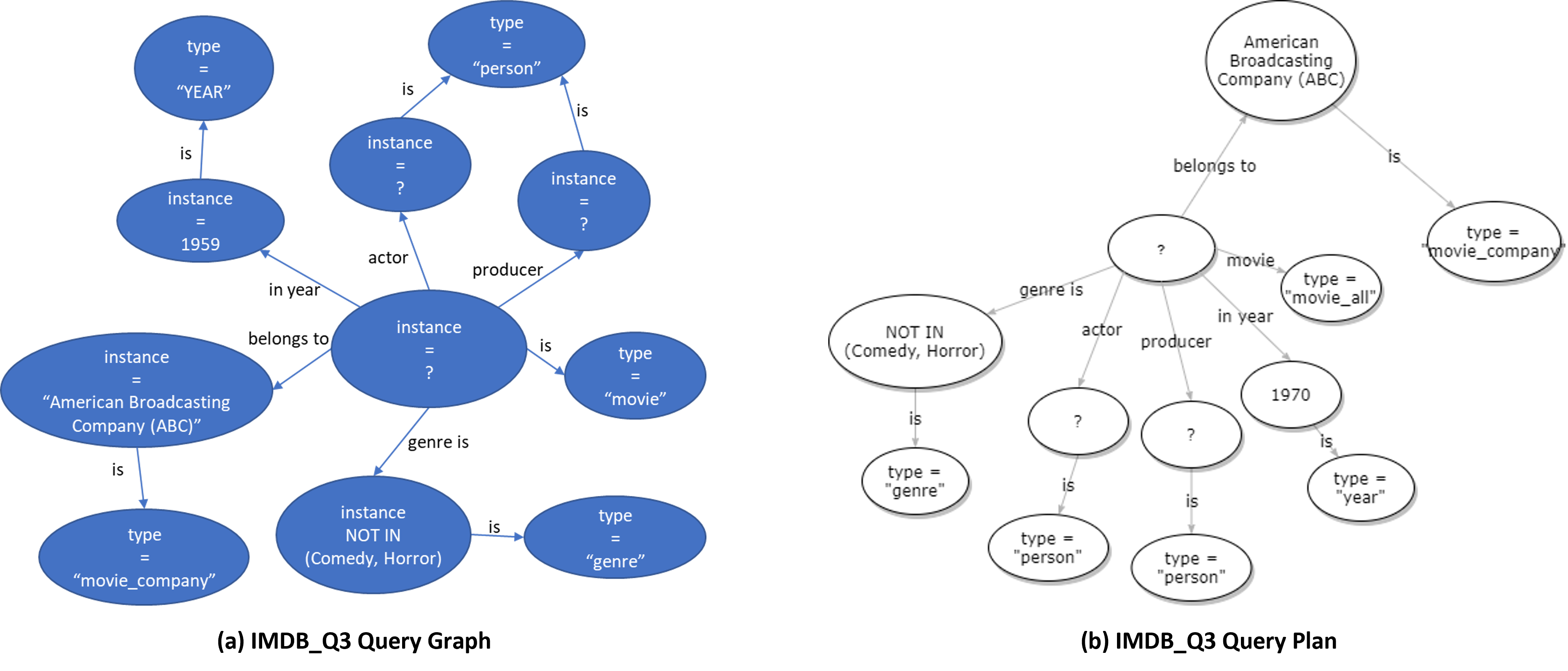}
%    \vspace{-5pt}
    \caption{Query Graph and Plan for IMDB\_Q3: \textit{Find movies along with their actors and producer information by company American Broadcasting Company (ABC) after year 1970 where the genre should not be comedy or horror.}}
    \label{fig:imdb-q3}
 %   \vspace{-20pt}
\end{figure*}

\textbf{Addressing Multi-Path Query Plans:}
Query plan nodes are numbered for this purpose. The issue with the multi-path query plan is handled by marking the node that continues to a different partition with the node id of the query plan node (which is unique). Figure \ref{fig:mark_node} shows the relative marking of node 14 for our previous example (Figure \ref{fig:mp_qp}). Once this partial answer continues to partition P2, we would know which query edge to expand further from the relative node id of the node and find correct results. 
In our case node $(14$:$4, B)$ in the partial answer will be further expanded for query edge $(14, B) \rightarrow (15, D)$.
% In our case node $(B, 14:4)$ in the partial answer will be further expanded for query edge $(B, 14) \rightarrow (D, 15)$.

\subsection{Input Query and Plan Representations}
Apart from the input graph, the query is given to our system as a graph. Of course, the query is a much smaller graph as computed to the input graph and has query-related decorations, such as wild cards, Boolean conditions, BETWEEN operator, and Boolean AND and OR specifications. Nodes and edges are labeled (including wild cards). This query is converted into a query plan with a root node and other nodes and edges taken from the query. Figures~\ref{fig:dblp-q3} and ~\ref{fig:imdb-q3} show two queries expressed as graphs along with their English description and one of the plans generated for each query. We have developed a query optimizer in ~\cite{DaWak/das2016plan} which uses a catalog generated for the input graph to perform a cost-based estimation of the plan. For this, we are using a plan that does not use a wildcard as the root node or top-level nodes as they match a large number of alternatives. We support the OR operator by combining the results of two plans as it corresponds to a UNION query. An example query is shown for each data set along with the plan.

These query results are further elaborated in Section~\ref{sec:experimental-analysis}. The plan is input as a text file (with extension .g) indicating node ids and their labels, followed by edges along with edge labels and nodes connected to them. Direction of edges can also be specified although we are using undirected plans here. Answer computation starts from the root nodes and generates all matches for the rest of the query plan. The search is constrained and does not generate node and edge labels not present in the plan.

\subsection{Correctness of the Proposed Approach}

For correctness, we need to find all exact matches of the query plan in the whole graph. As we have partitions of the graph, the result set should have answers from within a partition or across partitions. In our approach, we have addressed each case of answers spanning partitions. Also depending on query characteristics, we have addressed both \textit{single path} and \textit{multiple path} query plans.  Overall the correctness of the query processing is operationally ensured by i) Using Meta Information File (MIF) to correctly identify partitions that have start nodes (whether initial or continuation), ii) Updating Meta Information File to include partitions that have continuation or to remove partitions which have been processed, iii) Carrying intermediate results (instances with marked nodes) to the correct graph partition which will then load respective adjacency list for further expansion, and iv) Constructing partial answers and continuing expansion systematically from the last visited node.

\forceindent We have tested the correctness empirically (apart from the above operational correctness) as well by embedding graphs with known frequency and different structures in generated synthetic graphs using Subgen~\cite{net/subgen} and evaluating complete or partial queries of embedded graphs. It is not included due to space constraints. 
%%This is discussed in Section~\ref{sec:experimental-analysis}.

%% file: texFiles/mr-algorithms.tex
\section{Map/Reduce Architecture for Partitioning and Query Processing}
\label{sec:architecture-algorithms}

\begin{comment}
\begin{enumerate}
\item Start with selective expansion and how it is continued across partitions
\item block diagram of M/R
\item algorithms
\item what happens between iterations
\item stopping condition for all an top-k answers
\item how partial results are continued
\item operational correctness of algorithms
\end{enumerate}

\end{comment}

%\abhi{3/14/2022}{check}

\begin{figure}[H]
\vspace{-10pt}
    \centering
    \includegraphics[width=\columnwidth]{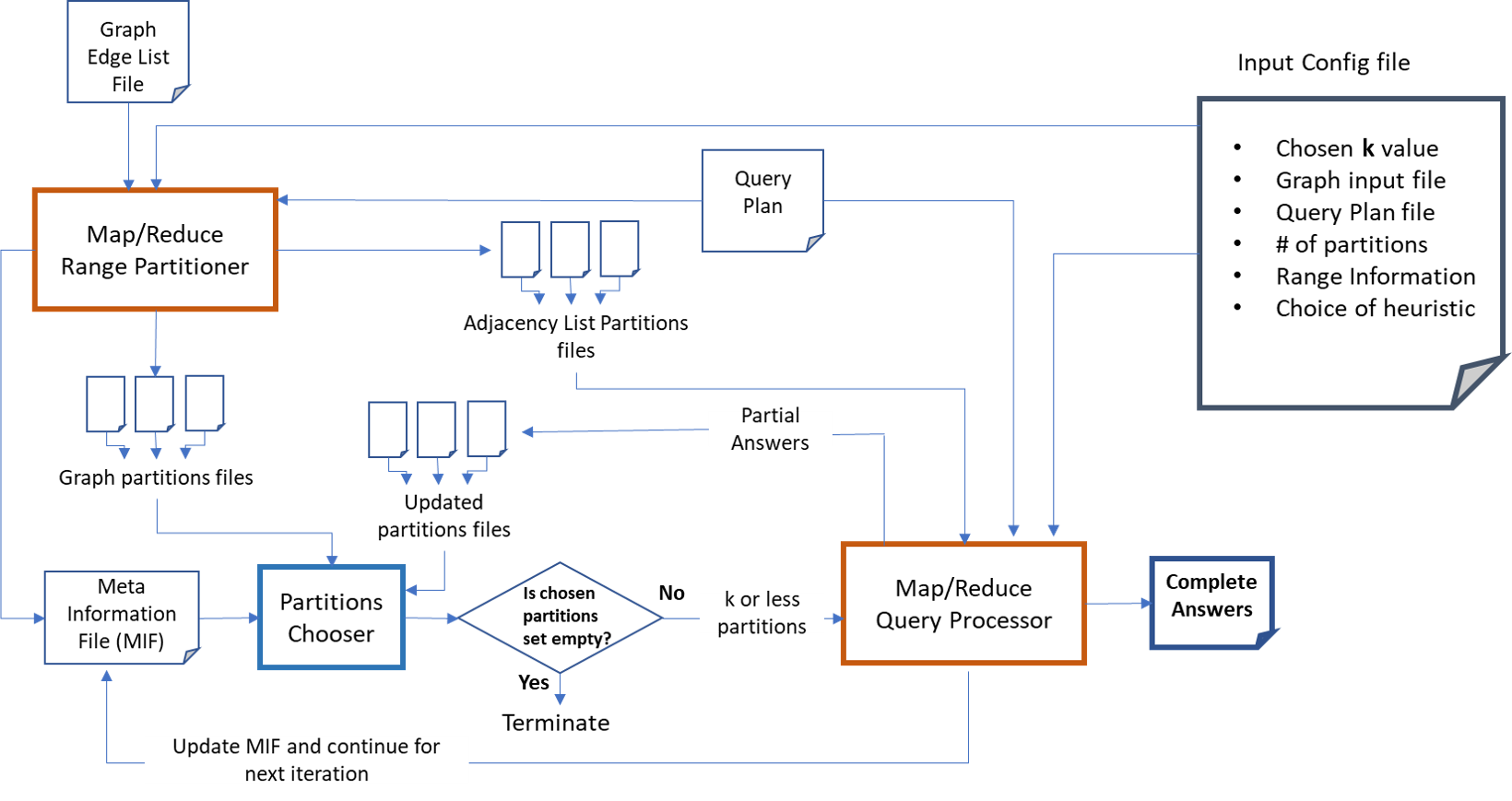}
    \caption{MR-GQP System Architecture}
    \label{fig:MR_GQP System Architecture}
    \vspace{-10pt}
\end{figure}

This section discusses the Map/Reduce algorithm for query processing. Input query, which is a graph (see Figures~\ref{fig:dblp-q3} and \ref{fig:imdb-q3}), is converted into a representation of edges and labels (including direction and wild cards) for processing. An input query is very small in terms of number of nodes and edges.  Figure \ref{fig:MR_GQP System Architecture} shows the overall query processing architecture using Map/Reduce. The input graph is range partitioned based on the resources available using the first pair of Map/Reduce processing. Meta Information File (MIF) is generated during partitioning, collecting query start and potential continuation nodes in each partition. Based on k, partitions are chosen using heuristics (discussed in Section~\ref{sec:overcoming-heuristics}.) Choosing partitions and processing a query continues until there are no more partitions to be processed. Map/Reduce query processor takes a query plan, adjacency list partition files and generates partial/complete answers as well as updates MIF for the next iteration. When finished all complete answers are written out.
%%SC:We will not discuss range partitioning using a Map/Reduce algorithm (refer Algorithm \ref{alg:range_partitioning}) due to space constraints. 

\begin{comment}
This section provides detailed description of the implementation of scalable query processing system based on partitioning and parallel processing. MR-QP developed as part of this paper uses Map/Reduce framework for a distributed processing to utilize parallel processing. Figure \ref{fig:MR_QP System Architecture} shows overall architecture of Map/Reduce based query processing system.
\end{comment}

\begin{figure}[h]
\begin{minipage}{0.6\columnwidth}
\centering
\vspace{-10pt}
\includegraphics[width=0.4\columnwidth]{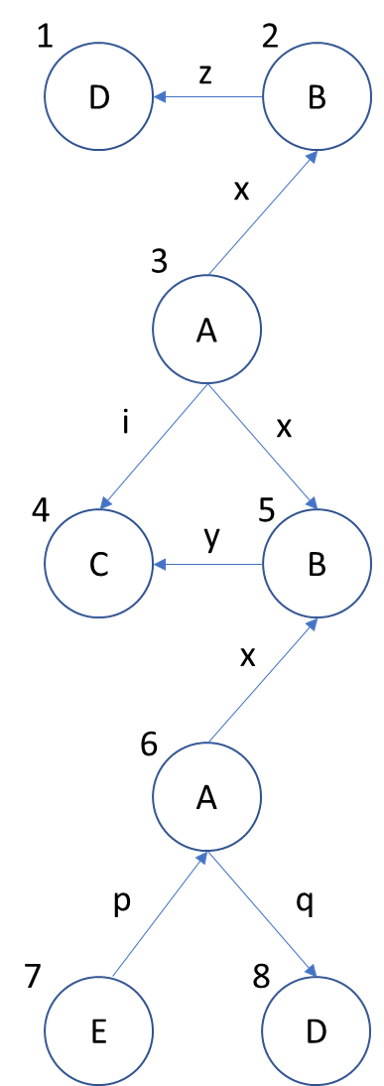}
\caption{Example Input Graph}
\vspace{-10pt}
\label{fig:input graph example}
\end{minipage}%
\centering
\begin{minipage}{0.4\columnwidth}
\begin{table}[H]
    \begin{tabular}{|p{1.5cm}|}
        \hline
            \textbf{Edges} \\
        \hline
        \hline
        (z,2,B,1,D) \\
        \hline
        (x,3,A,2,B) \\
        \hline
        (i,3,A,4,C) \\
        \hline
        (x,3,A,5,B) \\
        \hline
        (y,5,B,4,C) \\
        \hline
        (x,6,A,5,B) \\
        \hline
        (p,7,E,6,A) \\
        \hline
        (q,8,D,6,A) \\
        \hline
    \end{tabular}
    \caption{Graph Edge List Representation}
    \label{Graph Edge List Representation}
\end{table}
\end{minipage}%
\end{figure}

\noindent\textbf{Input Graph Representation:}
Our input graph is represented as a sequence of unordered edges (or 1-edge substructure). This makes it easier to distribute the graph partitions among many machines. Both undirected and directed graphs can be used for our approach but we have considered only directed graphs in our system as it has relations explicitly represented using directions. Table \ref{Graph Edge List Representation} shows the edge input representation of the graph in figure \ref{fig:input graph example}. Each edge is completely represented by a 5-element tuple \textit{$\langle $edge label, source vertex id, source vertex label, destination vertex id, destination vertex label $\rangle$}.

\forceindent Both partitioning and query processing are done using the Map/Reduce framework.
The driver program that runs Map/ Reduce jobs is implemented in Java. The configuration file with parameters is passed to the driver program.
Range partitioning is done in a single iteration of Map/Reduce, which generates graph partition files and adjacency list partition files. In addition, if query plans are available, partitioning generates an initial Meta Information File (MIF). MIF contains eligible partitions for query evaluation. After MIF is generated, our driver program applies the specified heuristic using the contents of MIF and chooses top-k partitions. When the chosen partitions set becomes empty, then query evaluation has been completed and we terminate the computation. Else we would have k or less partitions to process at a time, using which a Map/Reduce based query processing does query expansion on those partitions in parallel and outputs the complete answers to Final All Answers (FAA) file and writes the intermediate answers to the relevant graph partition files for further expansion. During the processing of partitions, MIF is updated and will be used by the driver program to select the next set of partitions to be loaded in the next iteration. This is continued until eligible partitions set in MIF are empty.

\begin{comment}

\begin{figure}[H]
    \centering
    \includegraphics[width=0.9\columnwidth]{figures/mr_partitioning.png}
    \caption{Range Partitioning using Map/Reduce}
    \label{fig:MR range partitioning}
\end{figure}

\end{comment}

\subsection{Range Partitioning using Map/Reduce}

%Figure \ref{fig:MR range partitioning} shows the range partitioning in Map/Reduce.
Algorithm \ref{alg:range_partitioning} gives the process of range partitioning using Map/ Reduce. %expands each mapper, partitioner and reducer class in details.

%\footnotesize
%\small
%{
\algnewcommand\algorithmicforeach{\textbf{for each}}
\algdef{S}[FOR]{ForEach}[1]{\algorithmicforeach\ #1\ \algorithmicdo}
\newcommand\sIf[2]{ \If{#1}#2\EndIf}

\begin{algorithm}[H]
\caption{Range Partitioning}
\label{alg:range_partitioning}
\vspace{2mm}
\footnotesize
\noindent \textbf{INPUT:} Graph edge list, Range info, Query Plan \\
\textbf{OUTPUT:} Graph partitions, Adjacency list partitions, MIF \\
\textbf{Class Mapper}
\begin{algorithmic}[1]
\Function{MAP}{key = line no, value = (one) edge instance}
    \State get the source vertex id($sVid$) and destination vertex id($dVid$) from edge
    \State emit(newKey = $sVid$, newValue = edge instance)
    \State emit(newKey = $dVid$, newValue = edge instance)
\EndFunction
\algstore{partitioning}
\end{algorithmic}
\textbf{Class Partitioner}
\begin{algorithmic}[1]
\algrestore{partitioning}
\Function{getPartition}{key = vertex id, value = one edge instance}
    \State reducerId = get partition id of vertex from Range Info
    \State return reducerId
\EndFunction
\algstore{partitioning}
\end{algorithmic}
\textbf{Class Reducer}
\begin{algorithmic}[1]
\algrestore{partitioning}
\Function{setup}{}
    \State $pId$ = get id of current reducer
    \State partitionRange = get range of partition $pId$ from Range Info
    \State edgeList = set to hold unique edges in current partition
    \State create new files for graph partition($pId$) and MIF
\EndFunction
\Function{REDUCE}{key = vertex id, valueList = list of adjacent edges}
    \State get vertex id($vId$) from key and vertex label($vLabel$) from valueList
    \If{$vLabel$ is Query Plan start node}
            \State update MIF with $\langle$ partition id = $pId$, label = $vLabel$, count = 1, \\ \qquad \qquad is start node = True $\rangle$
        \ElsIf {$vLabel$ is Query Plan intermediate node}
            \State update MIF with $\langle$ partition id = $pId$, label = $vLabel$, count = 1, \\ \qquad \qquad is start node = False $\rangle$
        \EndIf
    \State create initial substructure $\langle vId, vLabel \rangle$ and write out to graph partition($pId$)
    \State append every edge from $valueList$ to edgeList
    \State create adjacency list ($AL$) for vertex ($vId$)
    \State emit(newKey = $vId$, newValue = $AL$)
    % \ForEach {$ edge \in valueList $}
    %     % \State Get source vertex id(sVid), source vertex label(sVlabel) destination vertex \\ \qquad \quad id(dVid), destination vertex label(dVlabel) from value
    %     \State append edge to edgeList
    %     \State create single string($valueString$) from valueList
    %     \State emit(newKey = $vId$, newValue = $valueString$)
        
    % \EndFor
\EndFunction
\Function{cleanup}{}
    \State S = make sets for all vertices in current partition($pId$)
    \ForEach{edge $<sVid, dVid>$} in edgeList
        \If{sVid and dVid not in same set}
            \State union both sets of sVid and dVid
        \EndIf
    \EndFor
    \State count of connected components($CC$) = total unique sets in S
    \State update MIF with $\langle$ partition id = $pId$, connected components = $CC$ $\rangle$
\EndFunction
%\algstore{partitioning}
\end{algorithmic}
\end{algorithm}
%}

\normalsize

\textbf{Split and Shuffle Edges:}
Based on the number of mappers available, the Map/Reduce paradigm divides the input graph edge list into input splits and given as the input to the mapper. The map function then processes a single edge at a time and for both the vertex ids in the edge, it outputs the vertex id as the key and the edge instance as the value (line 1 to 5).

The shuffle phase sorts the mapper output based on vertex ids and groups all the records that have the same vertex id. As we want to divide output records based on range information, we use the custom partitioner that partitions keys with a custom hash function based on range information (lines 6 to 9). This will route all the output records with vertex id in the same range to the same reducer.

\textbf{Creating Graph Partition and Adjacency List Partition:}
Once the reducer gets the desired set of keys, we create graph partition files that contain initial substructures and adjacency list partitions that contain the adjacency list of all the vertices in the same partition. First, inside the setup of reducer, we create new graph partition file and local MIF (line 14). After that, the reduce function gets key as vertex id and value list as all adjacent edges from that vertex (line 16). For each vertex, we create initial substructure and output it to the corresponding graph partition file (line 25). Also, for each vertex, a single adjacency list is created. And finally, the reducer emits key as vertex id and value as adjacency list (line 27 and 28). The emitted keys that belong to the same partition are written to a single file creating the adjacency list partition.

\textbf{Collecting Meta Information and creating MIF:}
As we create initial substructures inside the reducer, we compare them to the query plan start node or intermediate nodes and if there is a match, we update the occurrences of those instances in the current partition in MIF (lines 18 to 24). Also, as the reducer gets all the edges within single partition, we collect all the edges (line 26) and in reducer cleanup function, using every edge in a partition, we find connected components using the find and union algorithm and count of connected components in that partition is written to MIF (lines 31 to 38). As each reducer processes separately, local MIF for each reducer will be created and then the driver program collects them to create global MIF.

\subsection{Query Processing using Map/Reduce}

%%SC: \begin{comment}

\begin{figure}[H]
    \centering
   \vspace{-15pt}
    \includegraphics[width=1\columnwidth]{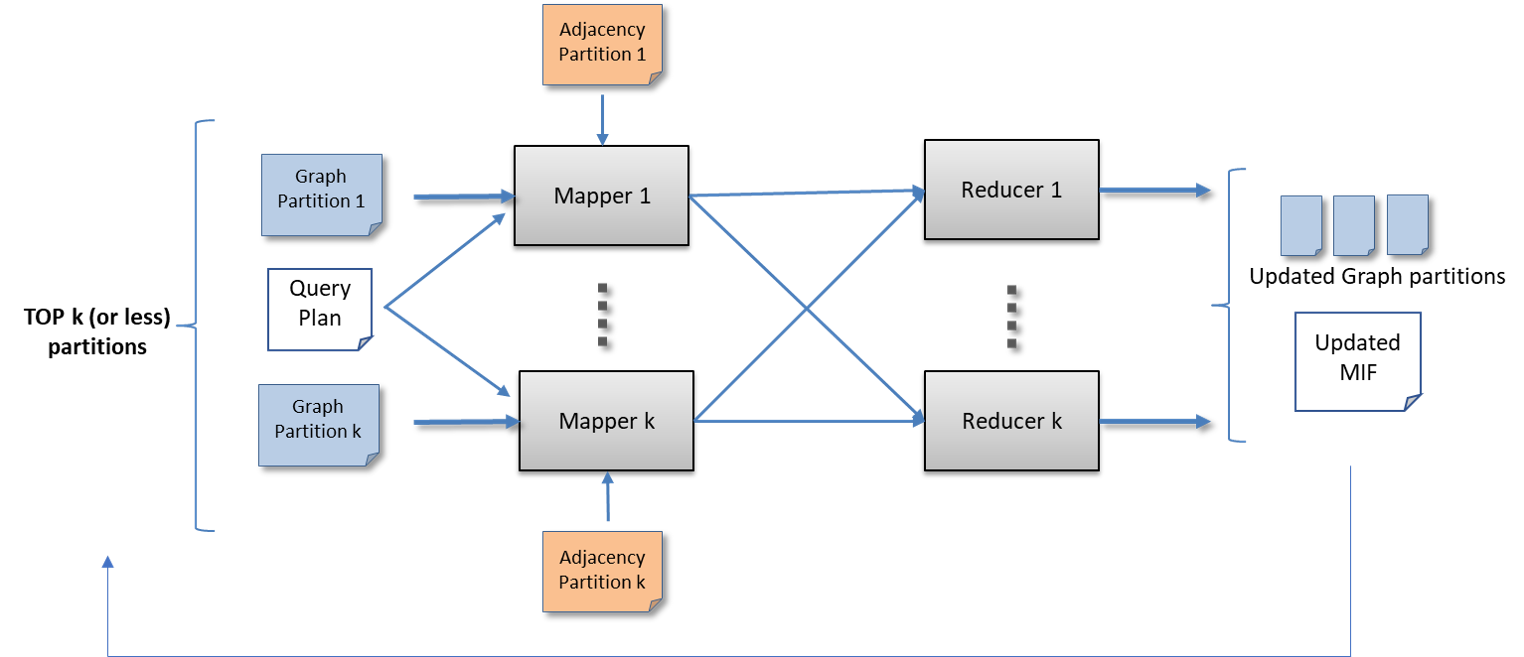}
    \caption{Query Processing using Map/Reduce}
    \label{fig:MR querying}
    \vspace{-12pt}
\end{figure}

%%SC: \end{comment}

%Figure \ref{fig:MR querying} shows the single iteration of Map/Reduce based query processing. 
%Algorithms for the mapper and reducer classes for query processing is presented in Algorithm \ref{alg:MR querying algorithm} and the Map/Reduce block diagram has been illustrated in Fig. \ref{fig:MR querying}. The top k or less graph partition files chosen from eligible partitions after applying heuristics are taken as input in each query processing iteration.

Algorithm \ref{alg:MR querying algorithm} presents the algorithms for the mapper and reducer classes for query processing. The Map/Reduce block diagram has been illustrated in Fig. \ref{fig:MR querying}. The top k or less graph partition files chosen from eligible partitions after applying heuristics are taken as input in each query processing iteration.

\noindent\textbf{Query expansion inside mapper:}
For query processing, input to each mapper is a graph partition that contains initial substructures or partial answer instances. In the setup function, the mapper loads the respective adjacency list partition and query plan (lines 3 and 4). We then create a set to hold unique completed answers from the current partition (line 5). 
The map function takes the key as the line number and value as the initial substructure or partial answer instance (line 7). Then we start query expansion for those instances that have query plan start node (line 8) and do a systematic expansion for every query edge present in the query plan and store the intermediate substructures in a set (lines 9 to 32). In the case of a partial answer instance, we will continue expansion only if the last visited node has been marked with the relative id of the current query node (line 16). To expand a node, we use the adjacency list of that node and according to the query plan, find all edges that match with the current query edge and expand the substructure with that edge (lines 18 and 19). \textit{If the newly expanded node belongs to a different partition}, then we mark that node relative to the last visited query node in the query plan and the mapper emits a new partition id as key and marks partial answer instance as the value (lines 25 and 26). Otherwise, if the expanded node is in the same partition, we create a canonical instance of the answer and store them in the answers set to remove duplicates (lines 20 to 22). We then continue with the next query edge from the query plan. When there is no more edge to expand from the query plan and no more partitions to be loaded, our answer set contains all complete answers. We collect them in the global answer set for that mapper (line 30). In the cleanup function, the complete answers are written in the Final All Answers (FAA) file (line 34).

\footnotesize
%\small

\begin{algorithm}
\footnotesize
\caption{Map/Reduce-Based Query Processing}
\label{alg:MR querying algorithm}
\vspace{2mm}
\textbf{INPUT:} Top k graph partitions and adjacency list partitions, Query Plan \\
\textbf{OUTPUT:} Updated graph partitions, Final answers file(FAA), Updated MIF
\textbf{Class Mapper}
\begin{algorithmic}[1]
\Function{setup}{}
    \State $currentPartitionId$ = partition Id from currently loaded graph partition
    \State adjMap = load corresponding adjacency partition
    \State load Query Plan file
    \State finalAnswers = Set to hold completed answers
\EndFunction
\Function{MAP}{key = line no, value = initial or partial answer instance}
    \If{value contains Query Plan start node}
        \State answerSet = Set to hold unique substructures after each query edge \\ \qquad \quad expansion
        \State add initial value substructure to answerSet
        \While{all the nodes and edges in the plan get visited}
            \For{$ans$ substructure in answerSet}
                \State get last visited node (v) from $ans$
                \State if node (v) has been marked and doesn't match with Query node \\ \qquad \qquad \qquad \qquad then continue for next query edge
                \State aL = get adjacency list of v from adjMap 
                \State expand the current node to the desired node by adding edge e\\ \qquad \qquad \qquad \qquad from aL based on the plan
                \If{$pId$ of expanded node matches currentPartitionId}
                    \State create canonical instance newAns
                    \State remove $ans$ from answerSet and add $newAns$ to answerSet

%%\algstore{querying}
%%\end{algorithmic}
%%\end{algorithm}
%%\begin{algorithm}[h]
%%\begin{algorithmic}[1]
%%\algrestore{querying}
%%\vspace{2mm}

                \Else
                    \State mark expanded node in $ans$ substructure with relative id of last visited query node
                    \State emit(newKey = $pId$ of expanded node, newValue = \\ \qquad \qquad \qquad \qquad expanded $ans$ substructure)
                \EndIf
            \EndFor
        \EndWhile
        \State finalAnswers $\displaystyle \cup$ answerSet
    \EndIf
\EndFunction
\Function{cleanup}{}
    \State write out finalAnswers to FAA file in HDFS
\EndFunction
\algstore{querying}
\end{algorithmic}
\textbf{Class Reducer}
\begin{algorithmic}[1]
\algrestore{querying}
\Function{reduce}{key = partition Id, valueList = intermediate results}
    \For{value in valueList}
        \State get the last expanded vertex (v) from the value
        \State update MIF file with $\langle$ partition id = key, label = null, count = 1, \\ \qquad \qquad is start node = True $\rangle$
        \State write out value to graph partition ($pId$) in HDFS
    \EndFor
\EndFunction
\end{algorithmic}
\end{algorithm}

\normalsize

\noindent \textbf{Aggregating partial answers and updating graph partitions in reducer:}
After mappers have been processed, the shuffle phase groups the output of mappers on the partition id and sends them to the corresponding reducer. The reducer processes key-value lists generated for each partition. The reduce function gets the key as partition id and value list containing partial answers in that partition (line 36). For each partial answer substructure instance present in the list, it updates the MIF to include continuing nodes in appropriate partitions (lines 37 to 40). According to partition id, it then loads the corresponding graph partition file from HDFS and write out partial answers in that file (line 41). %\texttt{The algorithm is given in the appendix and we will remove line numbers for CRC.}

%% file: texFiles/addressing-challenges-correctness.tex
%\section{Overcoming Challenges and Heuristics}
\section{Heuristics for Scheduling Partitions}
\label{sec:overcoming-heuristics}

Both meta-information (collected at the time of partitioning and in each iteration) as well as heuristics based on them are leveraged to find all answers efficiently. Heuristics are used for: \textbf{i)} \textit{identifying which k partitions need to be loaded in each iteration} and \textbf{ii)} \textit{loading the minimum number of required partitions (totally) to answer a query}. This is validated experimentally.

\subsection{\textbf{Heuristics Derived From Meta Information:}}

%Instead of using the meta information individually, combining them is likely to yield a better strategy for loading partitions when the k value is greater than 1. Hence, we have explored several combination heuristics which are discussed below.

%We further investigate combining each individual heuristics and validate the use of them to improve querying performance. The intuition behind this is that combining heuristics may yield better results than individual. We have formulated following 3 different heuristics:

In this paper, we have proposed three heuristics to determine an efficient order of loading the k partitions for query processing. Random loading of partitions for evaluating a query (\textbf{RANDOM} heuristic) is used as a baseline heuristic against which others are compared. Initially, we tried using individual metrics, such as \textbf{MIN\_SN} (\textbf{MAX\_SN}) or minimum (maximum) number of start nodes in a partition and \textbf{MIN\_CC} or minimum number of connected components.
%%SC: ~\cite{pgqp,DaWak/das2016plan}. 
As each of them individually does not capture the partition characteristic, we have combined them using a weighted formula and normalization to get better heuristics. \textit{In the experimental section, we compare both individual and composed heuristics to highlight the need for combining multiple individual heuristics to get better ones\footnote{For the composition of heuristics, based on the intuition, many compositions were tried before arriving at ones which better captured the essence of the partition characteristic.}.}

%%Three types of meta information for a partition have been considered - a) number of start nodes, b) number of connected components and c) number of intermediate nodes. Intuitively, individually, a) \textit{higher the start nodes in a partition}, higher number of query answers will be explored in the beginning and is likely to help reduce the number of partitions loaded if the answers span into other partitions, b) \textit{lower the number of connected components} will make an answer stay inside a partition more often and thereby reduce the total number of partitions used, and c) lower number of intermediate nodes in a partition is likely to lead to that partition getting processed multiple times. As part of the proposed heuristics, instead of considering these pieces of partition information in isolation, we have taken into consideration the correlation that exists among these factors.

\begin{enumerate}

\item \textbf{SN\_CC:} %In The intuition behind this is to not only use start nodes in each partition but also use connected components to choose top-k partitions. 
%AS: When this heuristic is used, partitions containing \textit{more start nodes} and \textit{less connected components} are given higher priority and loaded first. Based on this intuition, the SN\_CC heuristic is calculated as follows: 
Here the idea is to give higher loading/scheduling priority to the partitions that contain \textit{more number of start nodes} and \textit{less number of connected components} as they mostly will contain the maximum number of query answer instances. Based on this intuition, the SN\_CC heuristic is calculated as follows: 
%\vspace{-10pt}
\begin{equation}
H_{SN\_CC} = (e^{-0.04*CC}) * S
\end{equation}
%\vspace{-5pt}
where in a given partition, ~$S$ is the number of start nodes  and~$CC$ is the number of connected components. CC is used to dampen the effect of start nodes a bit. With this formula, the partition containing more start nodes and less connected components is given a higher heuristic value. But for partitions that have more connected components, the heuristic value decreases which is what we want.
\textit{Partitions with a higher value for this heuristic will be given higher priority for loading}.

\item \textbf{SN\_CC$_{2k}$:} 
%This also uses start nodes and connected components in each partition. It chooses  top 2k partitions first from the set of eligible partitions according to MAX\_SN heuristic and from that, it chooses top k partitions according to MIN\_CC. The intuition is that it utilizes both start nodes and connected components in each partition in choosing top k partitions.
From the set of eligible partitions, first, \textit{top 2k partitions} are chosen with \textit{maximum number of start nodes}. From this set, \textit{k partitions} are selected that have the \textit{least number of connected components}. Thus, this \textit{heuristic utilizes both start nodes} (MAX\_SN) \textit{and connected components} (MIN\_CC) \textit{in each partition to choose the top k partitions}.

\begin{comment}

\begin{table}[h!t]
\renewcommand{\arraystretch}{1}
\centering
%\vspace{-10pt}
    \begin{tabular}{|c|c|c|c|c|c|}
        \hline
         \textbf{\qquad k \qquad  } & \textbf{partitions loaded} & \textbf{\# of iterations} & \textbf{Load Factor} \\
        \hline
          1 & 5 & 5 & 1 \\
        \hline
         2 & 5 & 3 & 1 \\
        \hline
        \end{tabular}
\caption{$SN\_CC_{2k}$ heuristic performance}
\label{table:SN_CC_2k heuristic performance}
\vspace{-10pt}
\end{table}

Table \ref{table:SN_CC_2k heuristic performance} gives this heuristic performance for the example given in Figure \ref{fig:heuristic}. It also performs better than individual heuristics as for both k = 1 and k = 2, it loads only required partitions. Hence, load factor in this case is also 1.

\end{comment}

\item \textbf{SN\_IN:} In addition to start nodes in each partition, in this heuristic we consider the query plan intermediate nodes to choose top-k partitions. \textit{The partitions having large number of intermediate nodes compared to start nodes will have potentially more answers (though partial) and will be needed multiple times during processing}. The formula used for this heuristic is:
%\vspace{-15pt}

\begin{equation}
H_{SN\_IN} = ln(S+1)*\frac{S}{S+\frac{I}{2}}
\end{equation}
%\vspace{-5pt}

where for a given partition, ~$S$ is the number of start nodes and~$I$ is the number of intermediate nodes.
Thus, \textit{the higher the value of SN\_IN for a partition, the higher will be its loading priority}. It gives preference to the higher number of start nodes, but also takes into account the number of intermediate nodes. If the number of intermediate nodes is 0 or small, it does not have much effect. If the number of intermediate nodes increases, the heuristic value decreases. 
\end{enumerate}

Our experimental analysis compares all of the above heuristics to establish an efficiency pattern with respect to the choice of heuristic.

%% file: texFiles/experiments.tex
\section{Experimental Analysis}
\label{sec:experimental-analysis}

\begin{comment}

\begin{enumerate}
\item Design of experiments for testing challenges, evaluating heuristics, scalability, and efficiency
\item all experimental results and their discussion and comparison with analysis before
\item summary
\end{enumerate}

\end{comment}

%%\subsection{\textbf{Experimental Setup}}

% \sharma{}{start with environment description of comet and processors etc.}
% \harshit{}{addressed bellow}
\noindent\textbf{Processing Environment:}
All the experiments have been performed using Java with Hadoop on Comet cluster at SDSC (San Diego Supercomputer Center (\url{https://www.sdsc.edu/services/hpc/hpc_systems.html}). The Comet Cluster has 1944 nodes and each node has 24 cores
(built on two 12-core Intel Xeon E5 2.5 GHz processors) with 128 GB memory, and 320GB SSD for local scratch space.

%\begin{wraptable}{l}{0.48\columnwidth}
\begin{table}[H]
\renewcommand{\arraystretch}{1}
\centering
%\vspace{-5pt}
\begin{small}
    \begin{tabular}{|c|c|c|c|}
        \hline
         \textbf{Data Set} & \textbf{\#Partitions} & \textbf{\#Vertices} & \textbf{\#Edges}\\
        \hline
          DBLP & 120 & 3.3M & 13.5M\\
        \hline
          IMDB & 60 & 0.7M & 3.5M\\
         \hline
          Synthetic & 10 & 400K & 1.2M\\
        \hline
          DBLP\_small & 3 & 12K & 30K\\
        \hline
            IMDB\_small & 3 & 11K & 33K\\  
        \hline
        \end{tabular}
\caption{Data Set Description}
\label{table:datasets}
\end{small}
\vspace{-10pt}
\end{table}
%\end{wraptable}

\begin{table}[H]
\renewcommand{\arraystretch}{1.5}
\centering
%\vspace{-20pt}
%\begin{small}
 \footnotesize   
    \begin{tabular}{|p{1.2cm}|p{3.7cm}|p{1.5cm}|c|}
    
        \hline
         \textbf{Query Name} & \textbf{English Query } & \textbf{Key Operator} & \textbf{\#Ans}\\
        \hline
        \hline
        \multicolumn{4}{|c|}{\textbf{DBLP Queries}} \\
        \hline
          DBLP\_Q1 & Find papers along with their authors, conferences, and the publication year, which are published \textbf{between} 1990 \textbf{and} 1995 and the paper name includes the term ``Transaction Management in Multidatabase Systems" & Range  \hspace{10pt} (BETWEEN) & 18 \\
        \hline
        DBLP\_Q2 & Find papers along with their authors and publication year, which are published \textbf{after} 1995 in the conference VLDB or SIGMOD and the  paper name includes the term ``Transaction Management in Multidatabase Systems" & Comparison ($>$) & 45
        \\
         \hline
         DBLP\_Q3 & Find all conference papers along with the conference names that are published in 2005 or 2008 or 2009 by the author whose \textbf{first name starts with} `John'. (\textit{Query Graph and Plan in Figure \ref{fig:dblp-q3}}) & Special Character (\%) & 7685
        \\
        \hline
        \hline
        \multicolumn{4}{|c|}{\textbf{IMDB Queries}} \\
        \hline
         IMDB\_Q1 & Find movies produced by ``Adam Sandler" where the genres should be comedy \textbf{and} Sci-Fi but the year should not equal to 2000 & AND & 24 \\
        \hline
         IMDB\_Q2 & Find all actors (both Male and Female) who have worked in a movie owned by the company ``Paramount Pictures" where the genre of the movie is either Drama \textbf{or} Action. & Union (OR) & 6824 \\
        
         \hline
         IMDB\_Q3 & Find movies along with their actors and producer information by company American Broadcasting Company (ABC) \textbf{after} year 1970 where the genre should \textbf{not} be comedy or horror. (\textit{Query Graph and Plan in Figure \ref{fig:imdb-q3}}) & Combination ($>$, NOT) & 7252
        \\
        \hline
        \end{tabular}
\caption{DBLP and IMDB Queries}
\label{table:dblp-q}
%\end{small}
\vspace{-10pt}
\end{table}

\noindent \textbf{Data Set Description:} For this work \textit{labeled} graphs are required on which query processing can be performed. As shown in Table \ref{table:datasets}, we have used real-world (DBLP and IMDB) and synthetic graphs to show the effectiveness and scalability of our approach with the largest DBLP graph having 3+ million nodes and 13+ million edges.
%%AS: We do not describe the data sets as they are  familiar to our readers.
The DBLP data set contains information about authors, papers, conferences, and years as vertex labels and edges showing relationships among these vertex labels. Similarly, the IMDB data set contains information about movies, year, genre, person, gender, and movie company as vertex labels and edges represent the relationship among these vertex labels. The synthetic graph contains 2000 unique vertex labels and 4000 unique edge labels.

\begin{figure*}
    \centering
    %\vspace{-20pt}
    \subfloat[DBLP (BETWEEN query)\label{fig:Speedup DBLP (BETWEEN query)}]{
    \includegraphics[width=0.9\columnwidth]{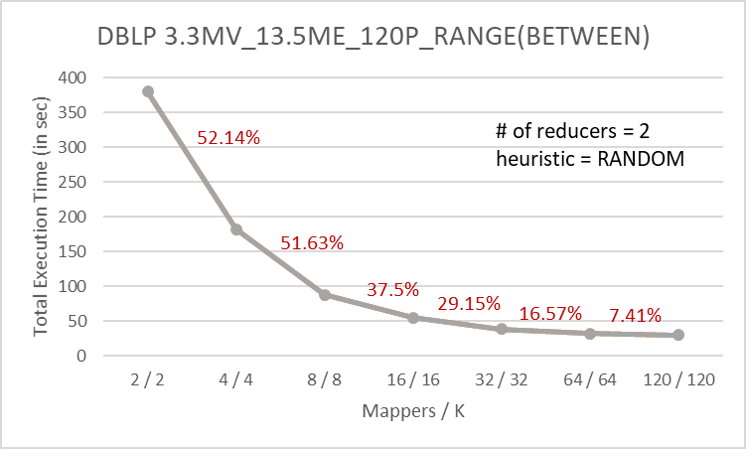}
    }\hfil
    \subfloat[DBLP (Wild Card (\%) query)\label{fig:Speedup: DBLP (Wild Card query}]{
    \includegraphics[width=0.9\columnwidth]{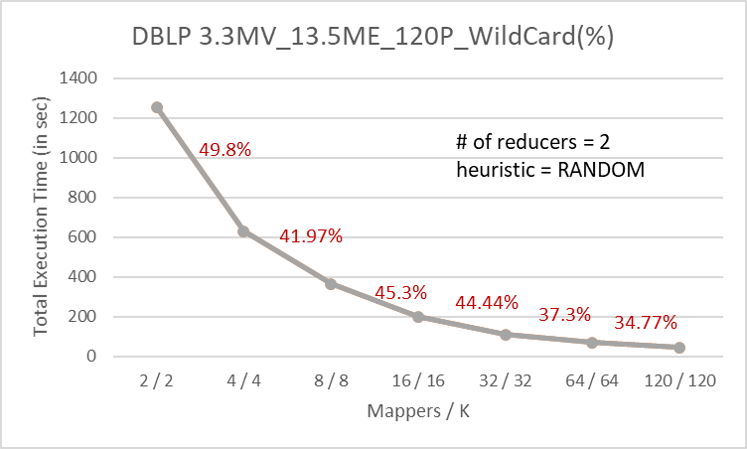}
    }

    \subfloat[IMDB (OR query)\label{fig:Speedup IMDB-ORQuery}]{
    \includegraphics[width=0.9\columnwidth]{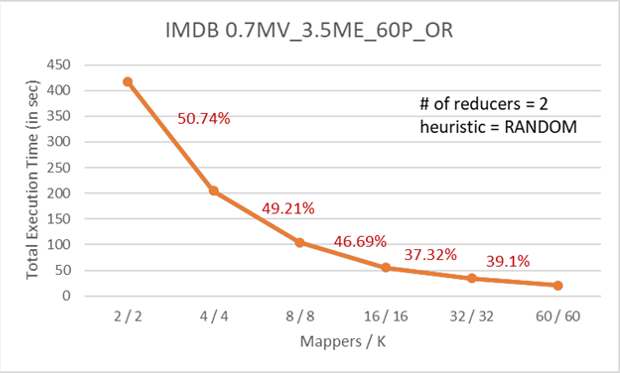}
    }\hfil
    \subfloat[IMDB (Combination query)\label{fig:Speedup: IMDB Combination}]{
    \includegraphics[width=0.9\columnwidth]{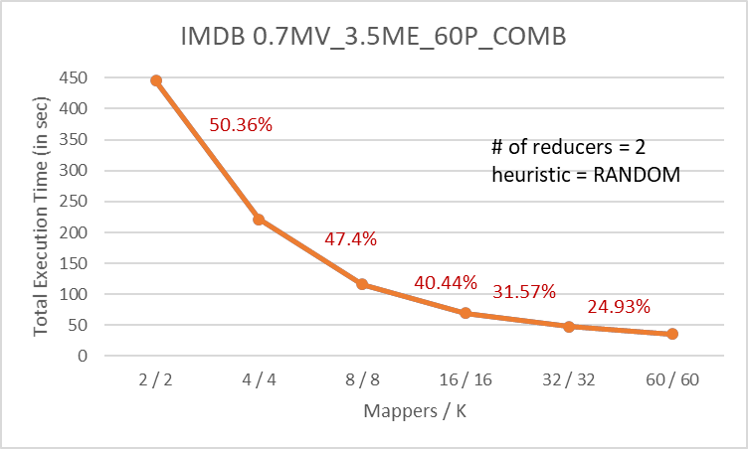}
    }
%    \vspace{-10pt}
    \caption{Response time Comparison of queries as k increases}
%    \vspace{-10pt}

\end{figure*}

\noindent \textbf{Correctness of Results:} For establishing the correctness of our algorithms, we have used an existing system, QP-Subdue, which is a non-partitioned, main-memory query processor for graph databases \cite{DaWak/das2016plan}.
%%SC: , msThesis/Goyal15}. 
However, QP-Subdue cannot handle larger graph sizes. Hence we have used synthetic data sets with embedded subgraphs as well as smaller sizes of both DBLP and IMDB graphs to verify the correctness, empirically, as shown in table \ref{table:datasets} for DBLP\_small and IMDB\_small.

\forceindent For the DBLP and IMDb graph data sets, we took queries having different characteristics (shown in Table \ref{table:dblp-q}). We have used queries with comparison operator, wild cards (\%), range queries (BETWEEN), AND query, OR query and combination of logical and comparison operators ($>$, NOT). Figure \ref{fig:dblp-q3} shows a sample query graph for the DBLP wild card query (DBLP\_Q3).

\subsection{\textbf{Results of  Partitioned Graph Querying}}

\noindent\textbf{Speedup with number of partitions processed in parallel in any iteration (k):}
% DaWAK Version
% The speedup achieved for one of the DBLP queries by increasing the value of k with respect to response time is shown in Figure ~\ref{fig:Speedup DBLP (BETWEEN query)}. The purpose of this experiment is to show change in response time with respect to resources used. As can be seen, for DBLP\_Q1 (BETWEEN operator query) there is \textbf{significant improvement in response time (starting with 54\%) until 64 partitions} (k) and tapers off after that. For the wild card query (DBLP\_Q3), shown in Figure~\ref{fig:Speedup: DBLP (Wild Card query}, even at \textbf{120 partitions there is significant improvement} as the number of answers are very large due to wildcard (see Table~\ref{table:dblp-q})!
The speedup achieved for different DBLP and IMDb queries by increasing the value of k with respect to response time is shown in Figure ~\ref{fig:Speedup DBLP (BETWEEN query)}. The purpose of this experiment is to show the change in response time with respect to resources used.

\forceindent As can be seen, for DBLP\_Q1 (BETWEEN operator query) there is \textbf{significant improvement in response time (starting with 52\%) until 64 partitions} (k) and tapers off after that. For the wild card query (DBLP\_Q3), shown in Figure~\ref{fig:Speedup: DBLP (Wild Card query}, even at \textbf{120 partitions there is significant improvement} as the number of answers is very large due to wildcard (see Table~\ref{table:dblp-q} last column)! Similar speedups are also observed for the IMDb queries, shown in Figure \ref{fig:Speedup IMDB-ORQuery} (IMDb\_Q2 - OR Query) and Figure \ref{fig:Speedup: IMDB Combination} (IMDb\_Q3 - Combination Query).

\begin{figure}[h]
    \centering
    \includegraphics[width=1\columnwidth]{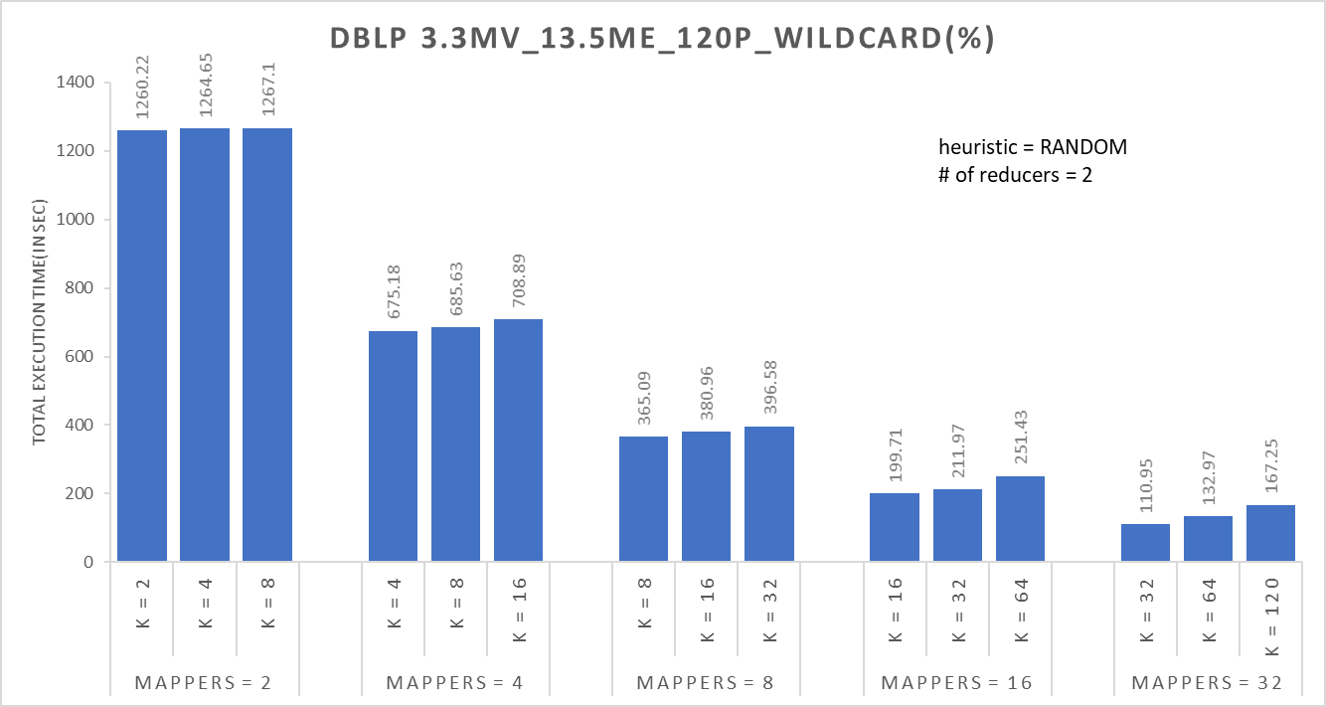}
    \caption{Effect of changing k for given mappers}
    \label{fig:Effect of changing k for given mappers}
  %  \vspace{-10pt}
\end{figure}

\vspace{5pt}
\noindent{\textbf{Effect of number of partitions for given mappers:}}
There are three parameters that affect the response time: i) the number of partitions (based on processor memory availability) or map tasks, ii) the number of physical mappers available for processing, and iii) the number of partitions loaded in each iteration. The total number of partitions used in Figure~\ref{fig:Effect of changing k for given mappers} is 120. Ideally number of partitions that can be used for each iteration needs to match the number of available mappers. When we cannot match \# of partitions in an iteration to the \# of mappers, multiple partitions have to be processed by each mapper in each iteration. In this experiment, we are interested in the effect on response time as we increase the k value (\# of partitions processed in each iteration) for a given number of mappers (fixed resources). We have used the Wild Card (\%) query here as it generates the maximum number of answers (DBLP\_Q3).

\forceindent Figure~\ref{fig:Effect of changing k for given mappers} shows two things: i) as the number of mappers (or resources) increases, response time goes down even if multiple partitions are processed due to an increase in k by each mapper (groups with different mapper values) and ii) when the number of mappers is held constant and k is increased making more partitions to be processed in each iteration, response time only increases slightly as multiple partitions are processed by the mappers ($2^{nd}$ and $3^{rd}$ bars compared to the $1^{st}$ bar in each set). Also, as the number of mappers increases, shuffle time increases which comes out as a bit higher value of response time as mapper size reaches 32. \textit{The takeaway is to match the number of partitions loaded in each iteration to the number of mappers available, if possible, for the best response time.}

\vspace{5pt}

\noindent \textbf{Effect of Heuristics:}
% DaWak version
%Instead of scheduling partitions for execution randomly, it would be better if we can reduce the total number of partitions loaded using heuristics. We have done these experiments to see the effect of each heuristic (discussed in Section~\ref{sec:overcoming-heuristics}) on query processing for all queries of DBLP graph. We only show the wild card query here. For initial analysis, we chose k = 1 and compared the response time for each heuristic. To see the effect of k, we increased k to 2, 4, 8 and 16 given that we have that many mappers available. Number of reducers for this set of experiments are fixed to 2.
Instead of scheduling partitions for execution randomly, it would be better if we can reduce the total number of partitions loaded using heuristics. We have done these experiments to see the effect of each heuristic (discussed in Section~\ref{sec:overcoming-heuristics}) on query processing for all queries of DBLP and IMDb graphs. We only show the wild card (DBLP\_Q3) and combination (IMDB\_Q3) queries due to space constraints. For the initial analysis, we chose k = 1 and compared the response time for each heuristic. To see the effect of k, we increased k to 2, 4, 8, and 16 given that we have that many mappers available. Number of reducers for this set of experiments are fixed to 2.

\begin{comment}
\begin{figure}
    \centering
    \includegraphics[width=1\columnwidth]{figures/heuristic_dblp_q1.png}
    \caption{Comparison of all Heuristics for DBLP (BETWEEN query)}
    \label{fig:Heuristics comparison: DBLP (BETWEEN query)}
\end{figure}

% \sharma{6/1/2020}{for some of these graphs, there are NO values on y-axis. why?}
% \harshit{}{addressed the same}

In case of BETWEEN query (Figure \ref{fig:Heuristics comparison: DBLP (BETWEEN query)}), for k = 1, SN\_CC is better than others as it gives less response time. SN\_CC$_{2k}$ and SN\_IN are second and third best heuristics respectively. On the other hand, RANDOM takes worst time among all. MIN\_SN takes worst time after RANDOM. For other k values, we see almost the same pattern. RANDOM takes worst time in all k values. MIN\_SN is second worst heuristic in each case. While SN\_CC is consistently better than others for all k values. SN\_IN and SN\_CC$_{2k}$ are the next top 2 heuristics.

\begin{figure}
    \centering
    \includegraphics[width=1\columnwidth]{figures/heuristic_dblp_q2.png}
    \caption{Heuristics comparison: DBLP ( Comparison $>$ query)}
    \label{fig:heuristic comparison DBLP comparison operator}
\end{figure}

For Comparison ($>$) operator (Figure \ref{fig:heuristic comparison DBLP comparison operator}), SN\_CC$_{2k}$ is consistently better than other heuristics for all k values. SN\_CC and SN\_IN comes under top 3 heuristics that gives us good response time. Here also RANDOM has worst performance for any k value. Moreover MIN\_SN is next worst heuristic.
\end{comment}

\begin{figure}
    \centering
    \subfloat[DBLP (Wild Card (\%) Query)\label{fig:heuristic comparison: DBLP Wild Card}]{
    \includegraphics[width=1\columnwidth]{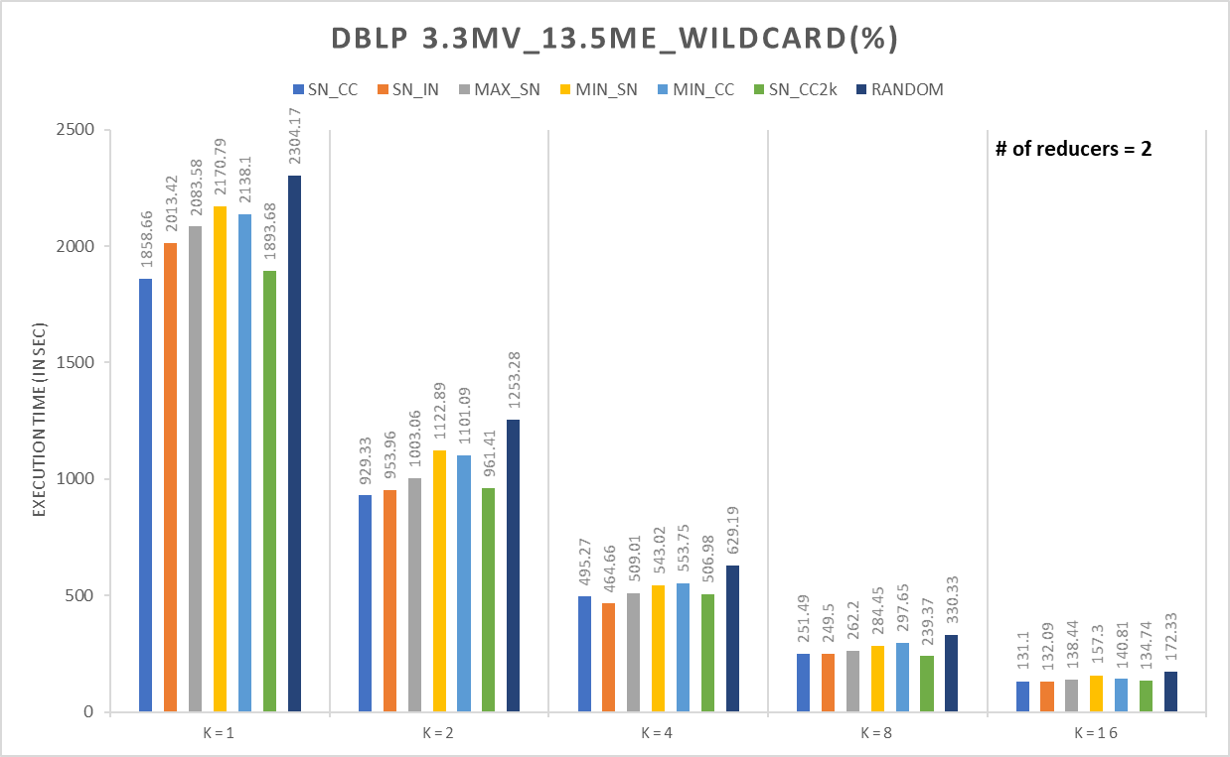}
    }\hfil
    \subfloat[IMDb (Combination Query)\label{fig:heuristiccomparison-IMDB-Combination}]{
    \includegraphics[width=1\columnwidth]{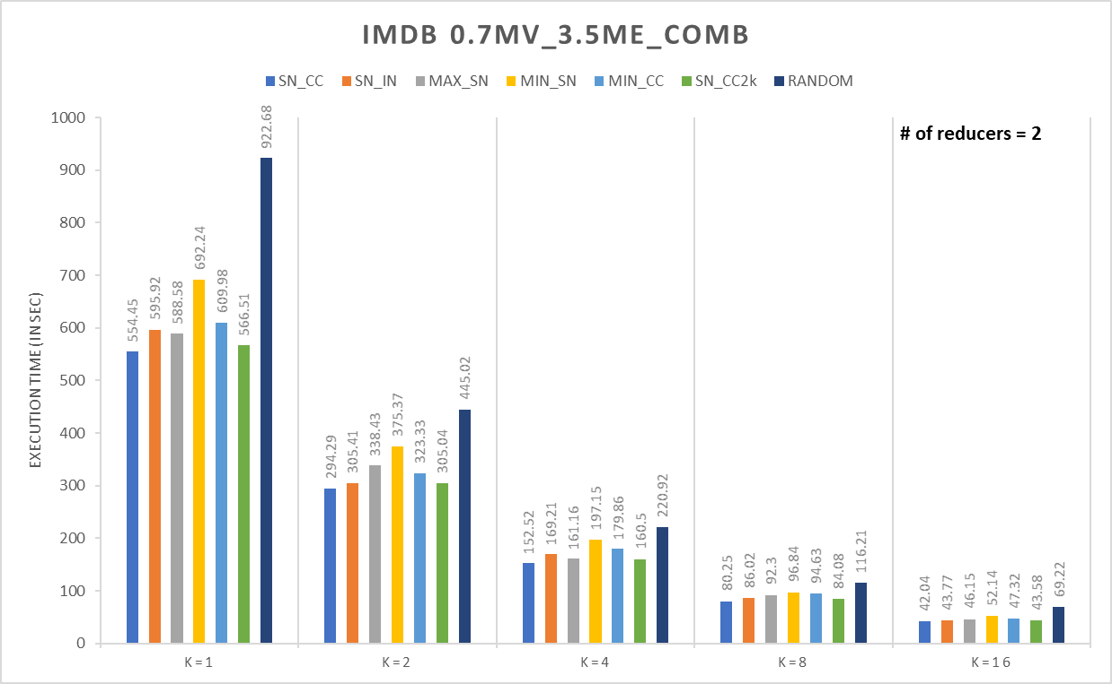}
    }
    \caption{Heuristics Comparison}
    \vspace{-10pt}

\end{figure}

%For wild card (\%) query (Figure \ref{fig:heuristic comparison: DBLP Wild Card}), the worst two heuristics are RANDOM and MIN\_SN. \textit{The heuristic that gives better performance than almost all others is SN\_CC$_{2k}$. The next top two heuristics for this query are SN\_CC and SN\_IN}. \\

\forceindent For the wild card (\%) query (Figure \ref{fig:heuristic comparison: DBLP Wild Card}) and the combination query (Figure \ref{fig:heuristiccomparison-IMDB-Combination}), 
it is observed in terms of execution time that the worst heuristics are RANDOM, MIN\_SN, MIN\_CC and MAX\_SN, while the \textit{best heuristics are SN\_CC$_{2k}$, SN\_CC and SN\_IN}. Thus, \textit{it has been showcased that the proposed composed heuristics perform much better than the individual heuristics}.

%\abhi{3/8/2022}{We have not discussed the RANDOM and MIN\_SN heuristics. Moreover, is there any justification for this result. Was the behavior expected? Can we add some takeaway from these experiments - \textit{All 3 (SS-IN, SS-CC, SS-CC2k) proposed heuristics perform better than the baseline (RANDOM) method across all the experiment settings}}. 
%%Similar things can be said about the load factor analysis.

\begin{comment}
\begin{figure}
    \centering
    \includegraphics[width=0.6\columnwidth]{figures/dblp_heuristics_comp.png}
    \caption{Comparing best and worst heuristics: DBLP}
    \label{fig:heuristic comparison best and worst}
\end{figure}
\end{comment}

%%\noindent\textbf{Comparing Heuristics:} To get the better understanding of the heuristics, we will compare top 3 and bottom 2 heuristics (5 in all) for all of the queries (figure \ref{fig:heuristic comparison best and worst}) for k = 1. As analysed above, our combined heuristics (SN\_CC, SN\_CC$_{2k}$ and SN\_IN) are coming in top 3. Further analysing these three heuristics, we can say that for BETWEEN and Comparison ($>$), we have lots of overlaps. While in case of Wild Card (\%), SN\_CC and SN\_CC$_{2k}$ are better than SN\_IN. \\

\subsection{Load Factor Analysis} 

Based on the chosen k value, we need to \textit{identify which k partitions need to be loaded in each iteration}. In the ideal case, we want to load the minimum number of required partitions to answer a query. %A required partition is one in which one or more number of the query plan nodes exist.
A required partition is one in which one or more query plan nodes exist.

\forceindent In processing a query over a partitioned graph, minimizing the total number of partitions loaded over all the iterations can be used as an \textit{efficiency measure}. %\textit{The optimal case is loading at most the number of required partitions only once.}
\textit{The optimal case is loading the number of required partitions only once. That is, the optimal load factor must be 1.}
%\abhi{3/8/2022}{check}

\begin{figure}[h]
    \centering
%    \vspace{-20pt}

    \includegraphics[width=1\columnwidth]{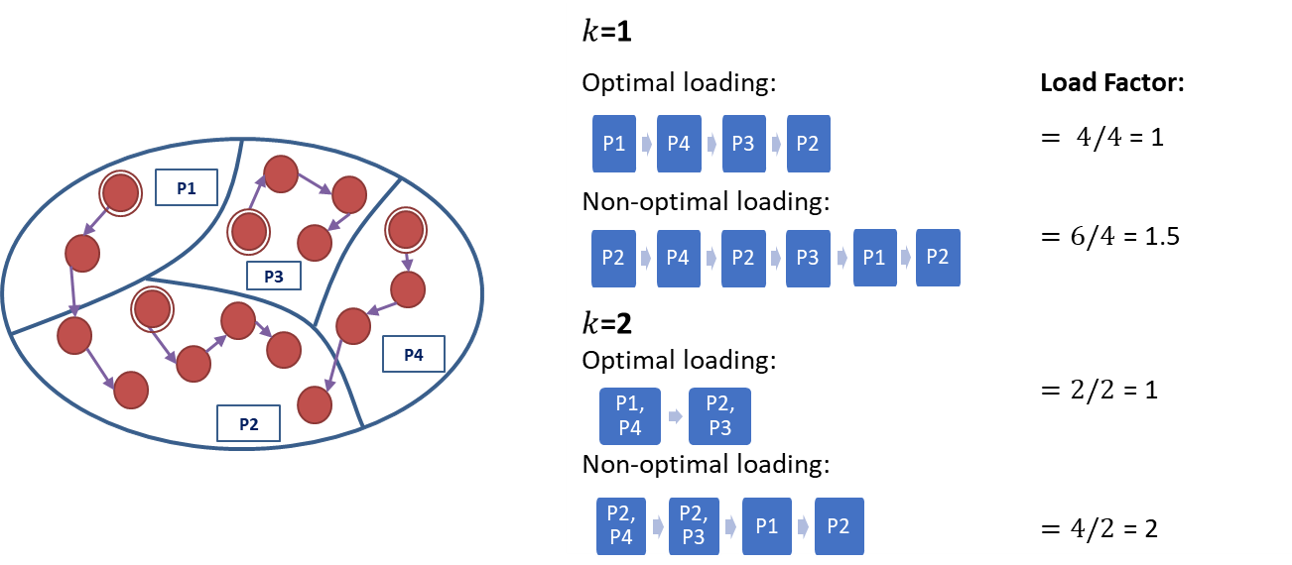}
    \caption{Load Factor Analysis: Optimal vs Non-optimal partitions loading}
    \label{fig:load_factor}
 %   \vspace{-20pt}
\end{figure}

\vspace{5pt}
\noindent\textbf{Load Factor:}
Apart from the number of partitions loaded, the query processing can take more iterations or fewer iterations depending upon the chosen value of k (and the partitions used in each iteration.) The ideal number of required iterations is when we load only the required partitions. We define `Load Factor' which measures the efficiency of query processing by comparing the number of iterations needed for the ideal case. The formula for load factor is defined as follows (illustrated in Figure \ref{fig:load_factor}):

\footnotesize
\begin{equation}
    \textit{Load Factor}
     = \frac{\textit{Total ~iterations ~taken ~for ~query ~processing}}{\textit{Required ~number ~of ~iterations}}
\end{equation}

\textit{where,}
\begin{equation}
    \textit{~Required ~number ~of ~iterations}
     = \lceil \frac{\textit{unique ~partitions ~that ~has ~answers}}{k} \rceil
\end{equation}

\normalsize

%For optimal loading as shown in figure \ref{fig:load_factor}, load factor is 1. This example also shows Non optimal sequencing, where some partitions are loaded more than once. As number of iterations also increases here, load factor for non-optimal loading increases. For k = 1, load factor is 1.5 whereas for k = 2, load factor is 2.

\forceindent The closer the value of load factor is to 1 the better the scheduling based on the chosen heuristic. We propose a suite of heuristics to identify which is better according to graph or query characteristics. If there are $p'$ required number of partitions and for a given k value, H(a) loads $p'$+a partitions and H(b) loads $p'$+b partitions, then we can say H(a) is better than H(b) if a $<$ b. In other words, \textit{the less the number of repeated partitions (or load factor), the better the heuristic. }
We further extend our heuristics comparison by taking the load factor for each heuristic into consideration.
\begin{comment}

\begin{figure}
    \centering
%    \vspace{-20pt}
    \includegraphics[width=1\columnwidth]{figures/load_factor_dblp.png}
    \caption{Load Factor analysis: DBLP Queries}
    \label{fig: Comparing load factor for all heuristics: DBLP)}
    \includegraphics[width=1\columnwidth]{figures/load_factor_imdb.png}
    \caption{Load Factor analysis: IMDb Queries}
    \label{fig: Comparing load factor for all heuristics: IMDB)}
%    \vspace{-20pt}
\end{figure}

\end{comment}

\begin{figure}
    \centering
    \subfloat[DBLP Queries\label{fig: Comparing load factor for all heuristics: DBLP)}]{
    \includegraphics[width=1\columnwidth]{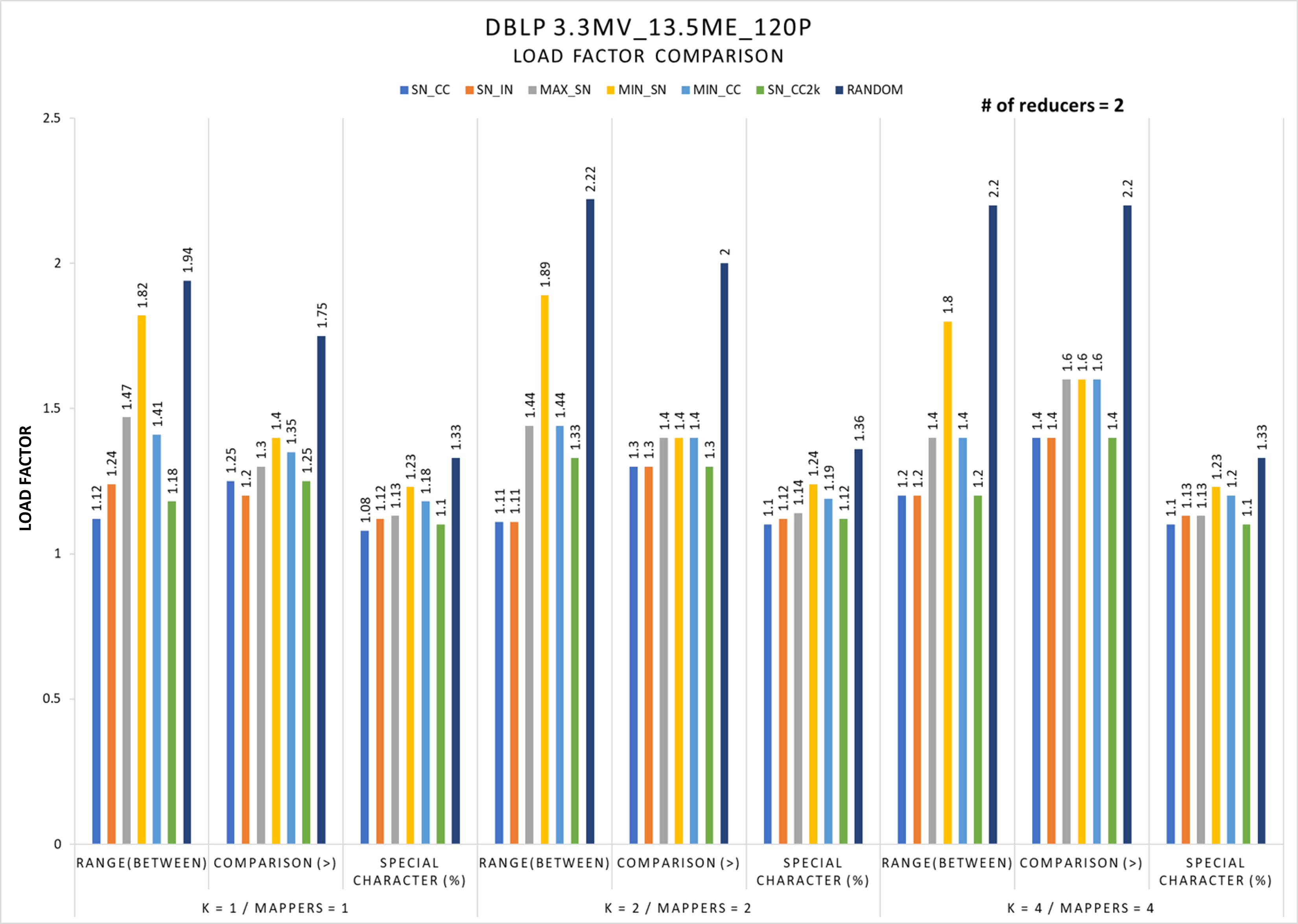}
    }\hfil
    \subfloat[IMDb Queries\label{fig: Comparing load factor for all heuristics: IMDB)}]{
    \includegraphics[width=1\columnwidth]{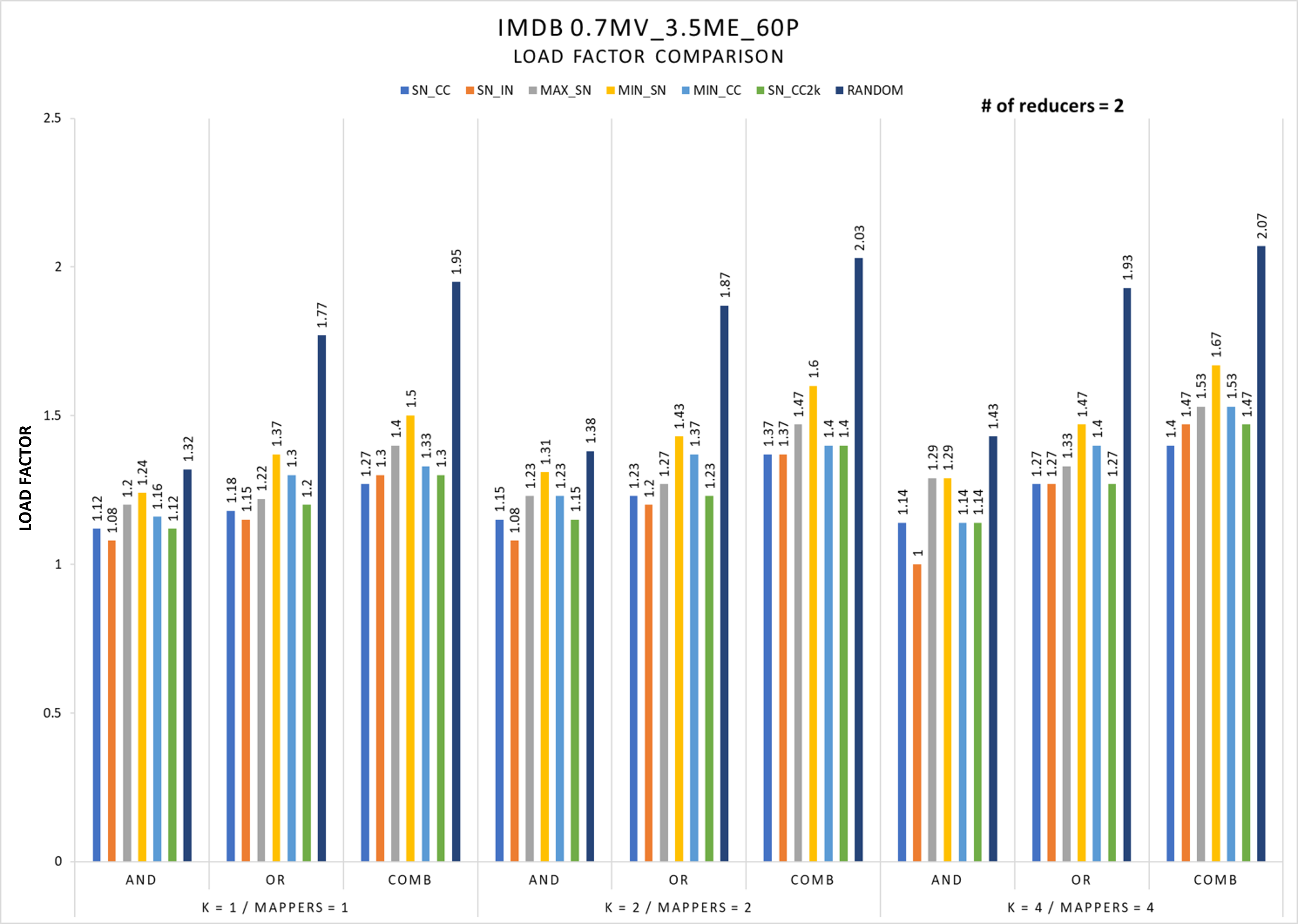}
    }
    \caption{Load Factor Analysis}
%    \vspace{-20pt}

\end{figure}

\forceindent Figures \ref{fig: Comparing load factor for all heuristics: DBLP)} and \ref{fig: Comparing load factor for all heuristics: IMDB)} show the load factor values for each heuristic for our given DBLP and IMDb queries, respectively. For instance in figure \ref{fig: Comparing load factor for all heuristics: DBLP)}, for k = 1, in case of BETWEEN query, SN\_CC has the lowest load factor value i.e. it requires less number of iterations and hence, less partitions are loaded for this heuristic. SN\_CC$_{2k}$ is the second best heuristic and SN\_IN is the third best heuristic in terms of Load Factor. Here, the load factor for the RANDOM heuristic is more than any other. Hence more partitions are loaded for the RANDOM heuristic, making it inefficient. The second worst is MIN\_SN. For all other DBLP and IMDb queries, the top 3 heuristics that have the lowest load factor values are SN\_CC, SN\_IN, and SN\_CC$_{2k}$. The presented analysis for load factor for all queries also matches with our previous heuristics comparison. We also see a similar pattern in load factor values for every value of k. \textit{In every scenario, the top 3 heuristics are SN\_CC, SN\_IN and SN\_CC$_{2k}$}.

\forceindent In summary, we have conducted extensive experiments on radically different data sets to present the results of our findings. Unfortunately, we are able to present few results due to space constraints. The queries that we have validated are quite expressive and are comparable to relational queries. We have also validated our approach for scalability of computation and verified the efficacy of proposed heuristics. We have shown speedup with respect to the number of mappers used (or k), compared all heuristics for all query types, and compared load factor for all heuristics. %\textbf{We have identified two heuristics (SN\_CC and SN\_CC$_{2k}$ that do better for all query types.}
\textbf{We have identified three heuristics (SN\_CC, SN\_CC$_{2k}$ and SN\_IN) that do better for all query types.}

%% file: texFiles/conclusions.tex
\section{Conclusions}
\label{sec:conclusions}

In this paper, we have proposed and demonstrated a scalable approach to efficient graph query processing using the Hadoop Map/Reduce framework that correctly computes the answers for different types of graph queries (including wild cards) efficiently. 
%%A number of challenges that arise due to partitioning of graphs and executing them independently are addressed in this paper. 
We have established the operational correctness of the algorithm as well in addition to experimental validation. 
%%Our approach has no restrictions on the query types as is commonly the case in other alternatives (e.g., only 3 entities and relationships). 
Experimental results on large real-world and synthetic data sets show the effectiveness and efficiency of this approach.

\forceindent Future work includes extending this framework for additional cost-based optimizations as is commonly done in the relational model. Also, techniques for partitioning the graph for a workload (of multiple queries) and processing \textit{multiple queries in parallel} (rather than sequentially and in addition to processing a single query in parallel) need to be developed. Whether partitioning changes make any difference for workloads with multiple queries needs to be investigated as well.

%% file: texFiles/acknowledgement.tex
\section*{Acknowledgments} 
For this work, Dr. Sharma Chakravarthy was  partly supported by NSF Grants CCF-1955798 and CNS-2120393.